\documentclass[useAMS,usenatbib]{mn2e}

\usepackage{graphicx}

\title[Radio halos in SZ-selected clusters of galaxies: the making of a halo?]{Radio halos in SZ-selected clusters of galaxies: the making of a halo? }

\author[A. Bonafede et al.]{A. Bonafede,$^1$\thanks{E-mail:
email@address}   H. Intema$^2$, M. Br\"uggen$^1$, F. Vazza$^1$, K. Basu$^3$, M. Sommer$^3$,  
 \newauthor  H. Ebeling$^4$, F. de Gasperin$^1$, H. J. A. R{\"o}ttgering$^5$,   R.J. van Weeren$^6$, R. Cassano$^7$.\\
$^1$ Hamburger Sternwarte, Universit\"at Hamburg, Gojenbergsweg 112, 21029, Hamburg, Germany. \\
$^2$ National Radio Astronomy Observatory, 1003 Lopezville Road, Socorro, NM 87801-0387, USA.\\
$^3$ Argelander Institut f\"ur Astronomie, Universit\"at Bonn, D-53121 Bonn, Germany.\\
$^4$ Institute for Astronomy, University of Hawaii, 2680 Woodlawn Drive, Honolulu, HI 96822, USA.\\
$^5$ Leiden Observatory, Leiden University, P.O. Box 9513, 2300 RA Leiden, the Netherlands.\\
$^6$ Harvard-Smithsonian Center for Astrophysics, 60 Garden Street, Cambridge, MA 02138, USA.\\
$^7$ INAF IRA, via Gobetti 101, 40129 Bologna, Italy.\\
}

\begin{document}

\date{Accepted Received...}

\pagerange{\pageref{firstpage}--\pageref{lastpage}} \pubyear{2002}

\maketitle

\label{firstpage}

\begin{abstract}
Radio halos are synchrotron radio sources detected in some massive galaxy clusters. Their Mpc-size indicates that (re)acceleration processes  are taking place in the host cluster. 
X-ray catalogues of galaxy clusters have been used in the past to search for radio halos and to understand their
connection with cluster-cluster mergers and with the thermal component of the intra-cluster medium.
More recently, the Sunyaev-Zel'dovich effect has been proven to be a better route to search for massive
clusters in a wider redshift range.
With the aim of discovering new radio halos and understanding their connection with cluster-cluster mergers,
we have selected from the Planck Early source catalog the most massive clusters, and 
we have observed with the Giant Metrewave Radio Telescope at 323 MHz those objects for which deep observations were not available.
We have discovered  new peculiar radio emission in three of the observed clusters finding:
(i)  a radio halo in the cluster  RXCJ0949.8+1708; (ii)  extended emission in Abell 1443 that we classify as a radio halo plus a radio relic, with a bright filament embedded in the radio halo;  (iii) low-power radio emission is found in CIZA J1938.3+5409 which is ten times below the radio - X-ray  correlation, and represents the first direct detection of the radio emission in the ``upper-limit" region of the radio - X-ray diagram.
We discuss the properties of these new radio halos in the framework of theoretical models for the radio emission.

\end{abstract}

\begin{keywords}
Galaxy clusters; non-thermal emission; particle acceleration; radio emission. Galaxy clusters: individual: Abell 1443, CIZAJ1938.3+5409, RXCJ0949.9+1708, RXCJ1354.6+7715
\end{keywords}

\section{Introduction}

Some galaxy clusters host diffuse radio emission which is not directly
connected to any of the cluster radio galaxies.  This emission fills
the inner Mpc of the host cluster (radio halo) or is located at the cluster periphery (radio relics), and is characterised by a low
surface brightness at $\nu \sim$ 1.4 GHz ($\sim$ 1
$\mu$Jy/arcsec$^2$), and steep radio spectrum\footnote{We define the
  radio spectrum as $ S( \nu) \propto \nu^{- \alpha}$.}  with $\alpha
>$1  \citep[see e.g. review by][]{Feretti12}.\\
Radio halos are produced by ultra-relativistic
electrons, with Lorentz factors $\gamma_{\rm{L}}\sim \rm{10^4}$,
spinning in large-scale $\mu$G magnetic fields.
The radio power at 1.4 GHz ($P_{1.4\, {\rm GHz}}$) correlates
with the X-ray luminosity ($L_X$) of the host cluster  \citep[e.g.][]{Liang2000}. Since $L_X$ is a proxy of the cluster mass, this correlation indicates that more massive systems host more powerful
radio halos. In addition, a radio bi-modality is found in the $L_X -
P_{1.4\, {\rm GHz}}$ plane \citep{Brunetti09}: clusters with the same X-ray
luminosity either host a radio halo, whose power follows the $L_X -
P_{1.4\, {\rm GHz}}$ correlation (referred as ``radio-on state"), or do not host radio emission (``radio-off state").

From a statistical point of view, not only the
power but also the occurrence of radio halos is known to increase with
 $L_X$, and hence with the mass of the host cluster 
\citep[e.g.][]{GTF99, Cuciti15}.

The physical mechanism
powering this emission is  debated \citep{2010ApJ...722..737K,Cassano10}.
However, a connection with the dynamical status of the cluster is established
\citep[e.g.][]{Buote01,Cassano10}. 

 Radio halos are likely formed
during cluster mergers, that inject in the intra-cluster medium (ICM) a considerable amount of energy. Part of the energy is dissipated through turbulent motions and 
shocks. Turbulence could re-accelerate an existing population of seed relativistic electrons and produce synchrotron emission \citep[e.g.][]{Brunetti01,Petrosian01}.

Alternatively, it has been proposed that radio halos result from inelastic  hadronic collisions between cosmic ray protons (CRp)  and thermal protons (hadronic models, \citealt{1980ApJ...239L..93D},  \citealt{1999APh....12..169B}). However, hadronic models  predict $\gamma$-ray emission that has not been observed by the {\it Fermi} Satellite \citep[e.g.][]{Ackermann14}. In some of the hadronic model formulations \citep[e.g.][]{2010ApJ...722..737K} they would require a different magnetic field in clusters with and without radio halos, which
is not observed \citep[e.g.][]{Bonafede11a,Govoni10}. In addition, hadronic models cannot explain radio halos with very steep spectral index ($\alpha>$1.5, e.g. \citealt{Brunetti08}).

Mixed hadronic and re-acceleration models have been also proposed (e.g. \citealt{BL11}, \citealt{Zandanel14}).\\
Although the pure hadronic models cannot explain the halos on the $L_X -P_{1.4\, {\rm GHz}}$correlation, 
Mpc-scale diffuse emission originating from hadronic interactions is expected in  the radio off-state clusters too.
 The first evidence for it has been found by stacking $\sim$ 100 images of massive clusters using the SUMMS survey (Sydney University 
Molonglo Sky Survey, \citealt{SUMMS}) by  \citet{Brown11}.
 According to \citet{BL11}, the electrons originating from hadronic collisions could produce radio emission that is a factor $\sim$ ten 
 below the $L_X -P_{1.4\, {\rm GHz}}$correlation. The same electrons could be the seeding particles re-accelerated by turbulence during a merger. 
 
We refer the reader to the review by  \citet{BJ14} for a detailed description of the proposed mechanisms.

Recently, radio halos have
been found in clusters with low X-ray luminosities
\citep{2011A&A...530L...5G}, and we have discovered the first example of radio halo in a 
massive cool-core cluster that does not show signs for a recent merger \citep{Bonafede14b}.
In addition, unexpected spectral features have been observed at low frequencies with early LOFAR 
observations, that have detected a steepening of the radio spectrum at low frequencies in Abell2256 \citep{vanweeren12}.
 Although the spectral measure is restricted to a small portion of the halo, it  indicates that more complex models, either inhomogeneous turbulence or mixed hadronic and leptonic models, have to be considered \citep{vanweeren12}.

 Before the advent of Sunyaev-Zel'dovich (SZ) large sky surveys, X-ray catalogs were the primary source
 of information to``hunt" for new radio halos.
 However, due to the dependence of the
X-ray luminosity on the square of the gas density, X-ray selected samples can be biased
towards low mass and cool-core systems. 
As noted by \citet{Basu12}, a better selection could be based on the SZ effect. 

The SZ signal integrated over the cluster angular extent measures the
total thermal energy of the gas and as such correlates closely with
the total cluster mass (see Planck collaboration, 2011, 2013 and references
therein). 
Indeed, also the SZ signal correlates strongly with the radio
power, as discovered by \citet{Basu12} and confirmed by \citet{SommerBasu,Cassano13}.
 After
mergers, the thermalisation of the gas in the ICM happens on larger
timescales than the boost in X-ray luminosity, while the SZ signal
will grow moderately and gradually \citep{SommerBasu}.  Hence, SZ surveys have the
potential of finding radio halos in late mergers, and in clusters that
are left out by X-ray selections. 
Recently, a radio bi-modality had been found in the
$P_{1.4\,GHz}-$SZ plane too, for clusters with $M_{500} \geq 5.5\times 10^{14} M_{\odot}$ \citep{Cassano13}.
\\

In this paper, we present a first sample of radio observations conducted with the Giant Meterwave Radio Telescope 
with the aim of searching for new radio halos in clusters that have a strong SZ signal.
We have selected from the Planck catalogues, published in 2011 and 2013 \citep{Planck11,Planck13}
the most massive clusters ($M_{500} > 6 \times 10^{14} M_{\odot}$) for which no radio observations were available to establish the presence of diffuse radio emission. 
In this paper we present the results from the first set of observations. The clusters presented in this work have been selected from the first Planck catalog \citep{Planck11}.

 The remainder of the paper is the following:
 in Sec. \ref{sec:radio} we describe the radio observations and the main steps of the data reduction. In Sec. \ref{sec:clusters} we analyse the
 results of the observations for each cluster. A discussion of the results is presented in Sec. \ref{discussion} and we conclude in Sec. \ref{sec:conclusions}.
 Throughout the paper we use a $\Lambda$CDM cosmological model with $H_0=$ 71 km s$^{-1}$ Mpc$^{-1}$, $\Omega_m=$0.27, $\Omega_{\Lambda}=$0.73.

\begin{figure*}
\vspace{100pt}
\begin{picture}(140,140)
\put(-200,-15){\includegraphics[width=10cm]{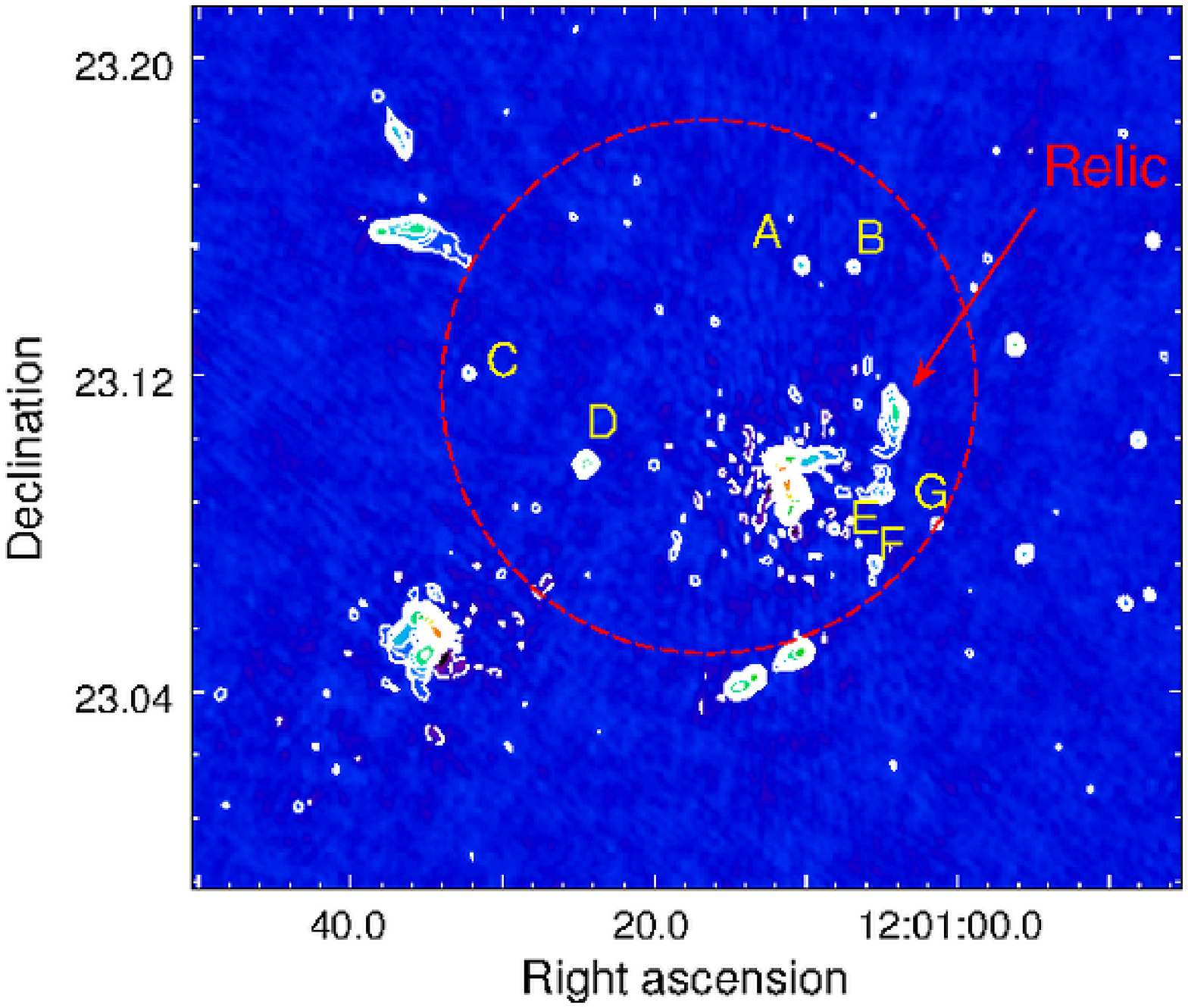}}
\put(70,10){\includegraphics[width=10.9cm]{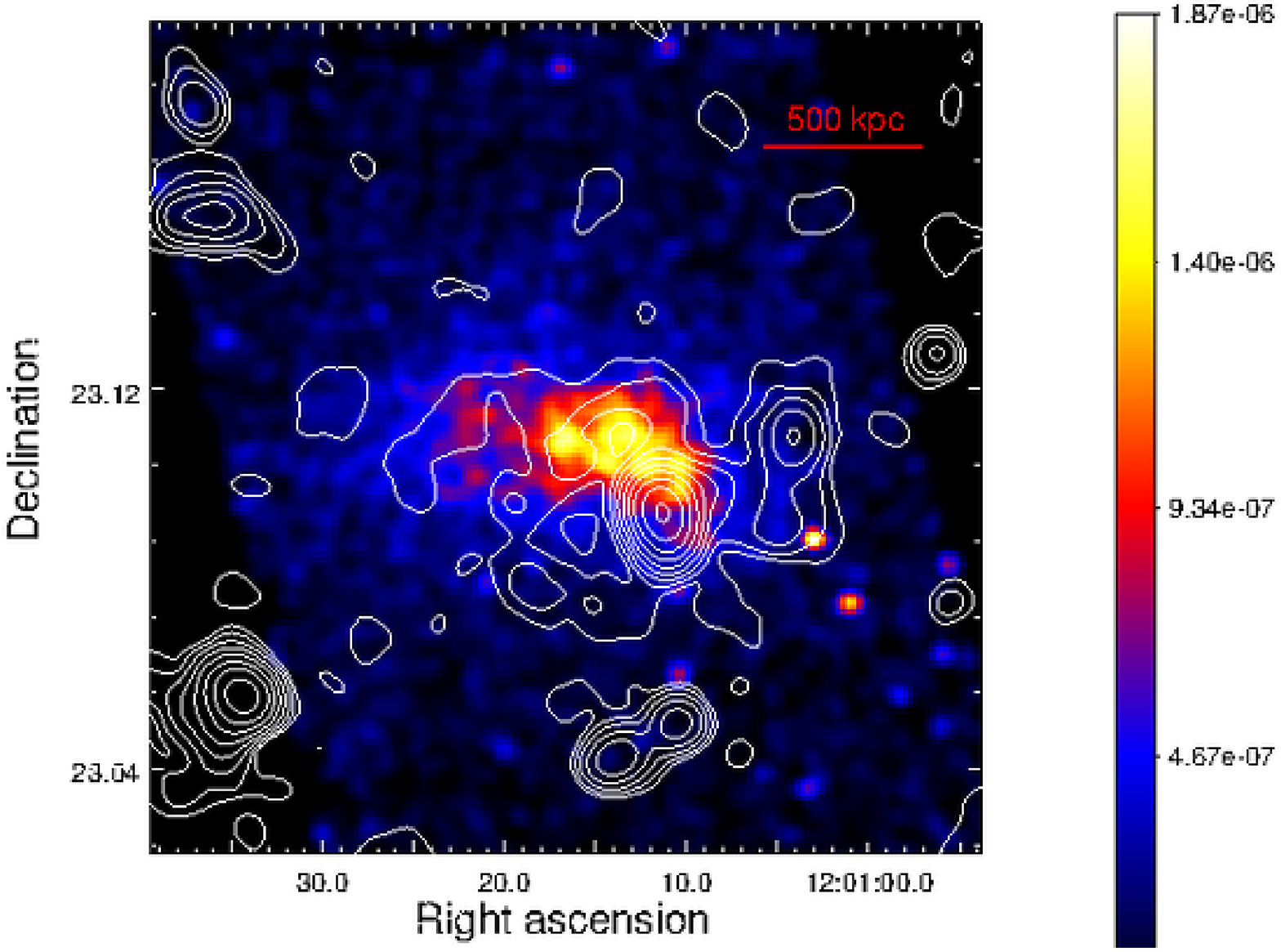}}

\end{picture}
\caption{Abell 1443: Left panel: HR image at 323 MHz in colours and iso-brightness contours. The beam is 9.8$" \times$8.1$"$. Contours start at 0.5 mJy/beam and are spaced by a factor 2. The first negative contour is dashed.
The red dashed circle is centred on the X-ray centre of the cluster and has a radius of 1 Mpc. The sources subtracted in the LR image are marked by letters (A to G). 
Right panel: Colours: Exposure-corrected X-ray emission of the cluster from {\it Chandra} in the 0.5 - 7 keV energy range. The colorbar has units of photons cm$^{-2}$ s$^{-1}$ pixel$^{-1}$. Radio iso-brightness contours from the radio LR image are overplotted. The beam is 27.4$" \times$26.7$"$. Contours start at 1.1 mJy/beam and are spaced by a factor 2. The first negative contour is dashed. }
\label{fig:A1443}
\end{figure*}

\begin{figure}
\centering
\includegraphics[width=7cm]{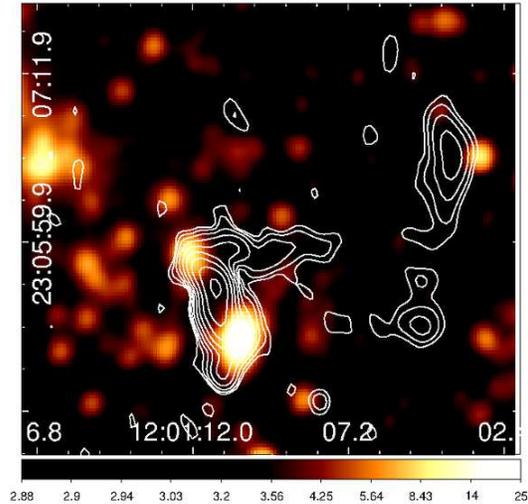}
\caption{{ Zoom into the $\Gamma$-shaped source and relic region of A1443. The contours are the same as in Fig \ref{fig:A1443}, left panel. Colours refer to the WISE 3.4 $\mu$m image. The colorbar is in magnitudes, not calibrated in terms of absolute surface brightness. }}
\label{fig:gammaSource}
\end{figure}

\begin{figure}
{\includegraphics[width=8cm]{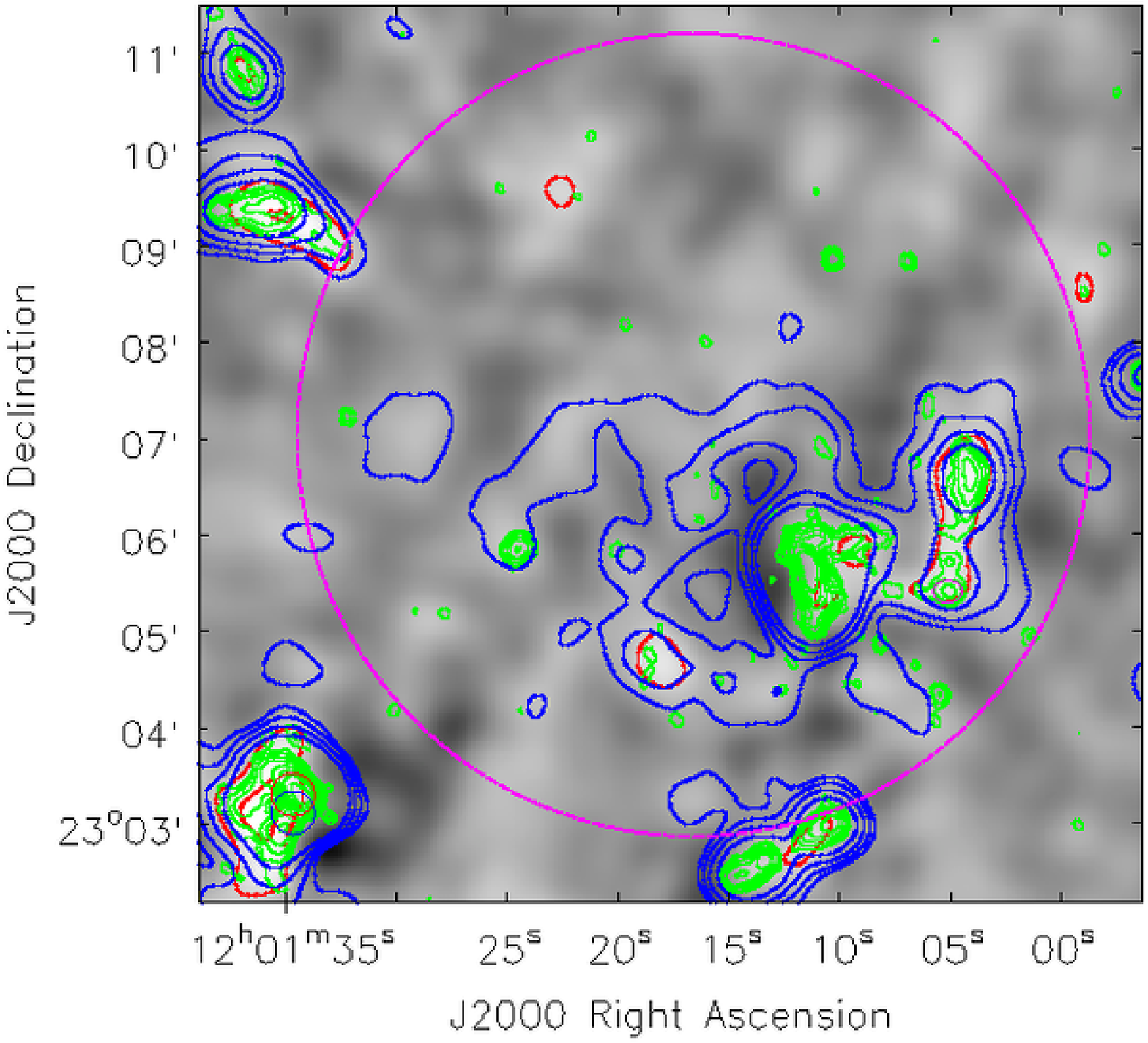}}
\caption{A1443: Grey colour scale: LR image of the cluster, blue contours are like in Fig. \ref{fig:A1443}, right panel. Green contours are like in Fig. \ref{fig:A1443}, left panel.  Red contours are the residual of the HR image convolved with a beam as the one of the LR image. Red contours are drawn at $\pm 3\sigma$. The dashed-circle is centred on the cluster centre and has a radius of one Mpc. The restoring beams are shown in the bottom left corner.}
\label{fig:A1443_residual}
\end{figure}

\begin{figure*}
\vspace{100pt}
\begin{picture}(130,130)
\put(-210,0){\includegraphics[width=11cm]{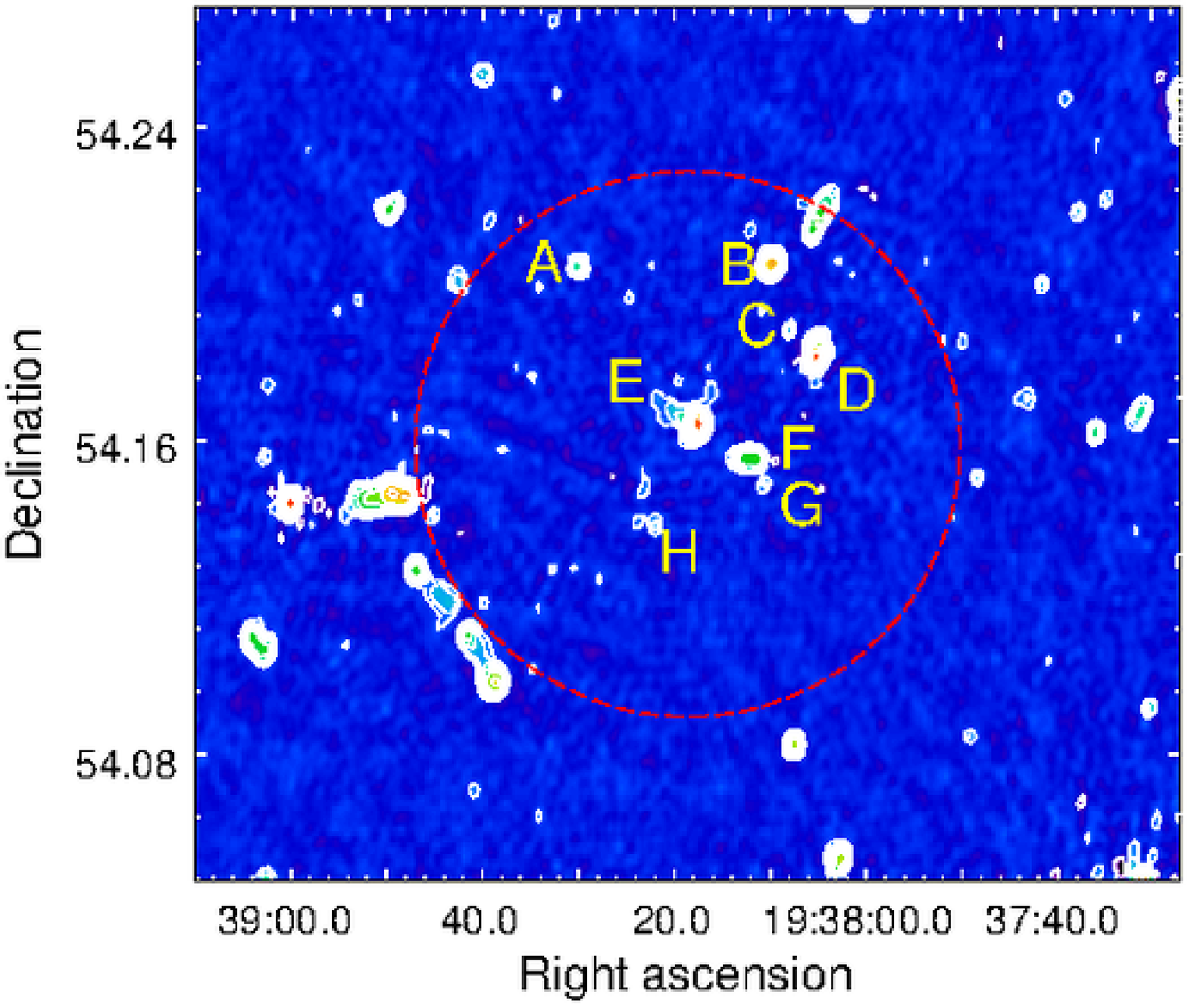}}
\put(55,10){\includegraphics[width=11cm]{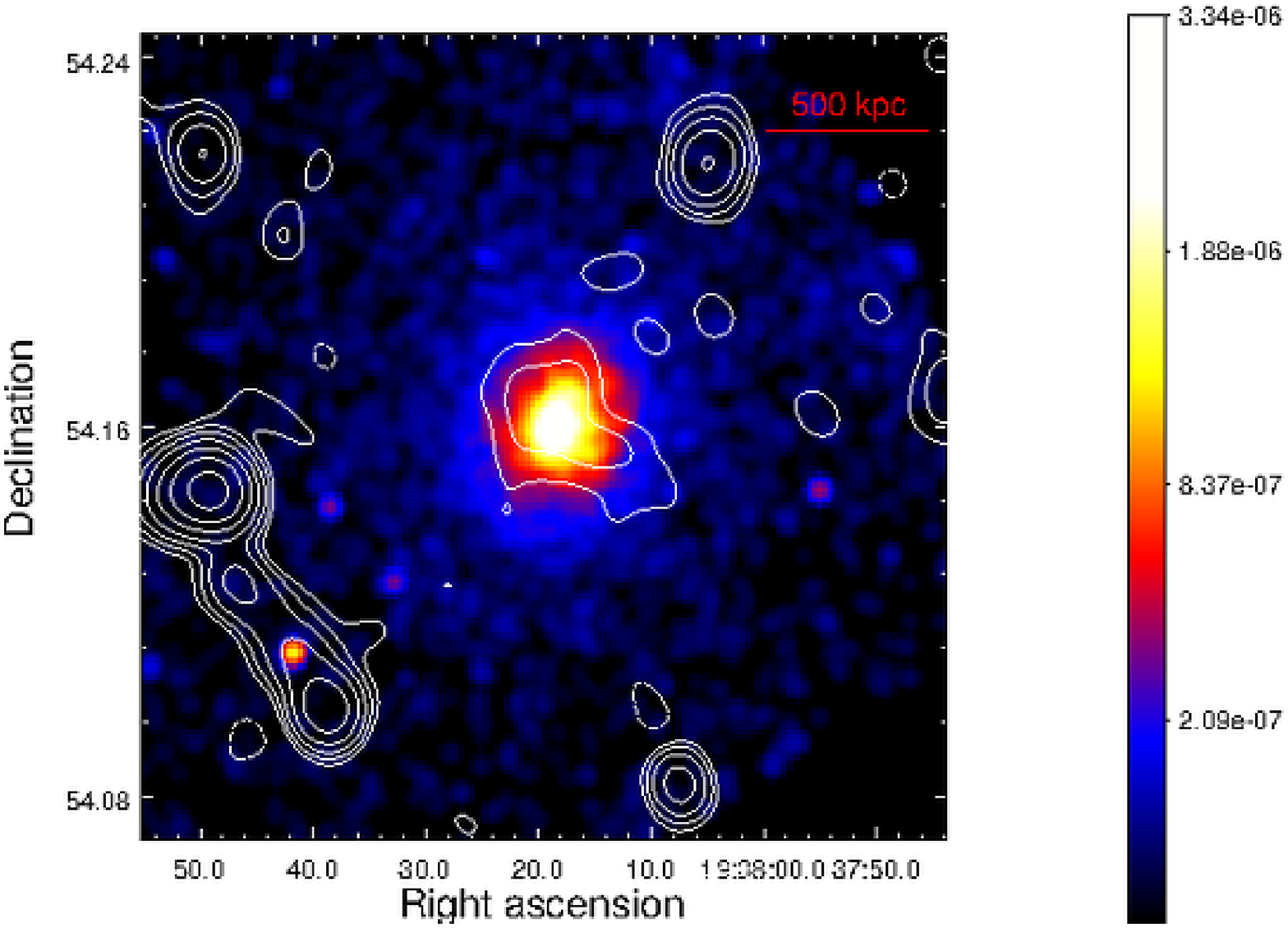}}
\end{picture}
\caption{CIZAJ1938.3+5409. Left panel: HR image at 323 MHz in colours and iso-brightness contours. The beam is 12.9$" \times$8.8$"$. Contours start at 0.3 mJy/beam and are spaced by a factor 2. The first negative contour is dashed.The red dashed circle is centred on the X-ray centre of the cluster and has a radius of 1 Mpc. The sources subtracted in the LR image are marked by letters (A to H). Right panel: X-ray emission of the cluster from {\it Chandra} in colours and iso-brightness contours from the radio LR image. The colorbar has units of photons cm$^{-2}$ s$^{-1}$ pixel$^{-1}$. The beam is 40.3$" \times$34.9$"$. Contours start at 0.7 mJy/beam and are spaced by a factor 2. The first negative contour is dashed.}
\label{fig:CIZAJ1938}
\end{figure*}

\begin{figure}
{\includegraphics[width=8.5cm]{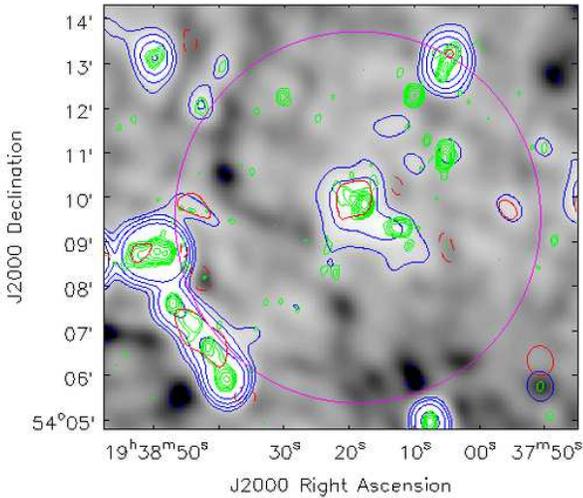}}
\caption{CIZAJ1938.3+5409: Grey colour scale: LR image of the cluster, blue contours are like in Fig. \ref{fig:CIZAJ1938}, right panel. Green contours are like in Fig. \ref{fig:CIZAJ1938}, left panel.  Red contours are the residual of the HR image convolved with a beam as the one of the LR image. Red contours are drawn at $\pm 3\sigma$. The dashed-circle is centred on the cluster centre and has a radius of one Mpc. The restoring beams are shown in the bottom right corner.}
\label{fig:CIZAJ1938_residual}
\end{figure}

\section{Radio observations and data reduction}
\label{sec:radio}
Observations were carried out at the Giant Meterwave Radio Telescope (GMRT) at 325 MHz, using a 33 MHz bandwidth subdivided into 256 channels and
8s integration time.
Depending on the position of the target, the sources 3C147, 3C48, or 3C286 were observed for 15 minutes at the beginning of the observing block, 
and used as absolute flux and  bandpass calibrator, adopting the  \citet{ScaifeHeald12} absolute  flux scale.
The absolute flux calibrators were also used to estimate  the instrumental contribution to the
antenna gains, which is needed for ionospheric calibration, and the instrumental phase information was used to correct the target field.
In Table \ref{tab:obs}  further details about the observations are listed.\\

In order to obtain a uniformly-reduced data sample, we have applied a semi-automated calibration procedure to the different datasets.
This approach minimises the  possibility of field-dependent effects introduced by a manual data reduction.

The main steps of the calibration procedure are outlined below, and are based on AIPS (Astronomical Image Processing System), SPAM \citep{SPAM} and Obit \citep{OBIT} tools. 
Strong Radio Frequency Interferences (RFI) was removed from the target field data by statistical outlier flagging tools. Much of the remaining low-level RFI 
was modelled and subtracted from the data using  Obit .
 After RFI removal, datasets have been averaged down to  24 channels, in order to speed up the imaging process and, at the same time,  avoid significant bandwidth smearing.\\
{ For the phase calibration, we started from a model derived from the Northern Vla Sky Survey (NVSS, \citealt{NVSS}) and WEsterbork Northern Sky Survey (WENSS, \citealt{wenss}) when available.
Using NVSS and WENSS, we have predcited the apparent flux of the sources at 323 MHz, using a simple spectral model fit and taking into account the GMRT primary beam.
Then, a cut in flux is applied to keep only the stronget sources. Typically, 20-30 sources are included in the first model.
Then, we proceeded with self-calibration loops.
  We decided not to use a phase calibrator, as the GMRT field of view is wide, and a non-negligible 
flux  is present in the field of the available  phase calibrators.
We note that using a phase calibrator  to bootstrap the flux from the absolute flux calibrator would have altered the flux scale, leading to higher errors in the flux measurements. 

The first imaging steps  have been performed using AIPS. In order to compensate for the non-complanarity of the array we used the wide-field imaging technique, decomposing  the GMRT field of 
view into $\sim$  80-100 smaller facets, depending on the field. 
 We performed few rounds of cleaning and self-calibration, and inspected the residual visibilities for a more accurate removal of low-level RFI. 
In order to correct for ionospheric effects, leading to direction-dependent phase errors, we applied SPAM calibration and imaging to the target field.
The presence of strong sources in the field-of-view enables one to derive directional-dependent gains for each of them (similarly to the peeling technique) and to  use these gains to fit a phase-screen over the entire field of view.
After ionospheric corrections, the last step of the calibration procedure consists of subtracting all the sources outside the central 15$'$ from the target centre to facilitate the imaging steps.}

Processed data were  imaged in CASA (the Common Astronomy Software Applications package), and some more cycles of self-calibration have been performed.
We have imaged the datasets at high resolution, using the Briggs weighting scheme (robust parameter equal to 0 in the CASA definition), and excluding the baselines shorter than 1.1 -0.9 k$\lambda$, depending on the cluster's redshift,  to filter out the possible emission on Mpc-scale.
High resolution images (HR images) have been used to identify the radio radio galaxies in the cluster and to subtract their emission from the uv-data.
Specifically: we have selected the clean components corresponding to the radio galaxies detected in the HR images within 1 Mpc from
the cluster centre, and we have subtracted the clean components from the visibilities. We have used these new datasets, where the sources' emission had been subtracted, to search for diffuse radio emission in the clusters.
To increase the brightness sensitivity to the diffuse emission, we have produced low resolution images (LR images)
tapering down the baselines longer than 4 k$\lambda$ and using the Briggs weighting scheme (robust parameter equal to 0.3 in CASA definition). 

The final images of the full field-of-view centred on the target were  corrected for the primary beam response.
We estimate that the residual amplitude errors are of the order of 10\% at 325 MHz in line with values
reported for GMRT observation at this frequency  \citep[e.g.][]{2009A&A...506.1083V,Intema11,Bonafede12}.\\

\begin{table*}
\label{tab:obs}
 \centering
   \caption{Radio observations.}
  \begin{tabular}{c c c c c c c c c }
  \hline
Planck name &  alt. Cluster name     &   Obs. date &   RA and DEC &  z & scale &Flux cal  & $\sigma_{HR}$  & $\sigma_{LR}$ \\ 
                  &					&			&   \scriptsize{ h m s, d $'$ $''$}   &     & \scriptsize{ kpc/$''$} &                       &   \scriptsize{mJy/beam}  &   \scriptsize{mJy/beam}            \\
 PSZ1 G229.70+77.97&	Abell 1443                  & 27/01/2013  & 12 01 27.7    +23 05 17.9       &   0.27   &   	 4.101 	& 3C147        &  0.13  & 0.30           \\
PSZ1 G086.47+15.31 &	CIZAJ1938.3+5409    & 26/01/2013  & 19 38 18.6   +54 09 33      &    0.26   &  3.990  & 3C48-3C286          &   0.10 & 0.15         \\
PSZ1 G216.60+47.00&   RXCJ0949.8+1708    & 25/01/2013  &	09 49 51.7 +17 07 08   	& 0.38   &   5.175 & 3C147                          &   0.15  & 0.20        \\
PSZ1 G118.46+39.31&    RXCJ1354.6+7715     & 26/01/2013  & 13 54 37.8   +77 15 35      &   0.40    &  5.343 & 3C286                    &    0.14 & 0.30        \\
\hline
\hline
\multicolumn{9}{l}{\scriptsize  Col. 1: Planck name, Col. 2: Other name from previous catalogues;  Col. 3: Observation date  }\\
\multicolumn{9}{l}{\scriptsize  Col. 4: Right Ascension and Declination of the target; Col. 5: target  redshift; Col. 6: Source used to set the absolute flux scale; }\\
\multicolumn{9}{l}{\scriptsize  Col. 7: Angular to linear scale; Col. 8:  rms noise in the high resolution image; Col 9: rms noise of the low-resolution image.}
\end{tabular}
\end{table*}

\begin{table*}
\label{tab:sources}
 \centering
   \caption{Radio sources}
  \begin{tabular}{c c c c c c c }
  \hline         
Planck name &  Other name   & Detection            & Flux density   & Radio power at 323 MHz & LAS $\times$ SAS   & LLS $\times$ SLS \\    
			&                                &			            &   mJy		     &  $ \rm{W \,Hz}^{-1}$& $'' \times ''$    & kpc$\times$kpc  \\
	
PSZ1 G229.70+77.97  &	Abell 1443                  &		Whole emission   & 580$\pm$59 &     1.4$\pm$0.1 $\times 10^{26}$          & 330$\times$230   & 1350$\times$940 \\
											 &		&Halo(?)   				    &  74.0 $\pm$7.6  &     1.7$\pm$0.2 $\times 10^{25}  $         &270$\times$230  & 1100$\times$940 \\
											 &		&Relic(?)   				    & 57.0$\pm$5.8  &       1.3$\pm$0.1 $\times 10^{25}   $        &160$\times$65    & 660$\times$270 \\
											 &		&$\Gamma$-shaped source    &450$\pm$45 &       1.1$\pm$0.1 $\times 10^{26}$      &  100$\times$60  & 410$\times$250 \\
PSZ1 G086.47+15.31 &	CIZAJ1938.3+5409    & 		Halo             			    & 11.0$\pm$1.2   &       2.4$\pm$0.3 $\times 10^{24}$        & 180$\times$100 &  720$\times$400\\
PSZ1 G216.60+47.00 &	RXCJ 0949.8+1708  &           Halo  					     &  21.0$\pm$2.2 &      1.1$\pm$0.1 $\times 10^{25}$             & 200$\times$80 & 1040$\times$400 \\

\hline
\hline
\multicolumn{7}{l}{\scriptsize  Col. 1: Planck name, Col. 2: Other name from previous catalogues;  Col. 3: Extended source detected in the cluster; }\\
\multicolumn{7}{l}{\scriptsize  Col. 4: Flux density of the detected source, or upper limit to it, Col. 5: Radio power of the detected source assuming a spectral index $\alpha=1.3$ for the \it{k}-correction;}\\ 
 \multicolumn{7}{l}{\scriptsize  Col. 6: Largest and smallest angular scale of the source;  Col. 7: Largest and smallest linear scale of the source in the adopted cosmological model. }
\end{tabular}
\end{table*}

\section{Individual clusters}
\label{sec:clusters}

\begin{figure*}
\vspace{100pt}
\begin{picture}(130,130)
\put(-210,-10){\includegraphics[width=11cm]{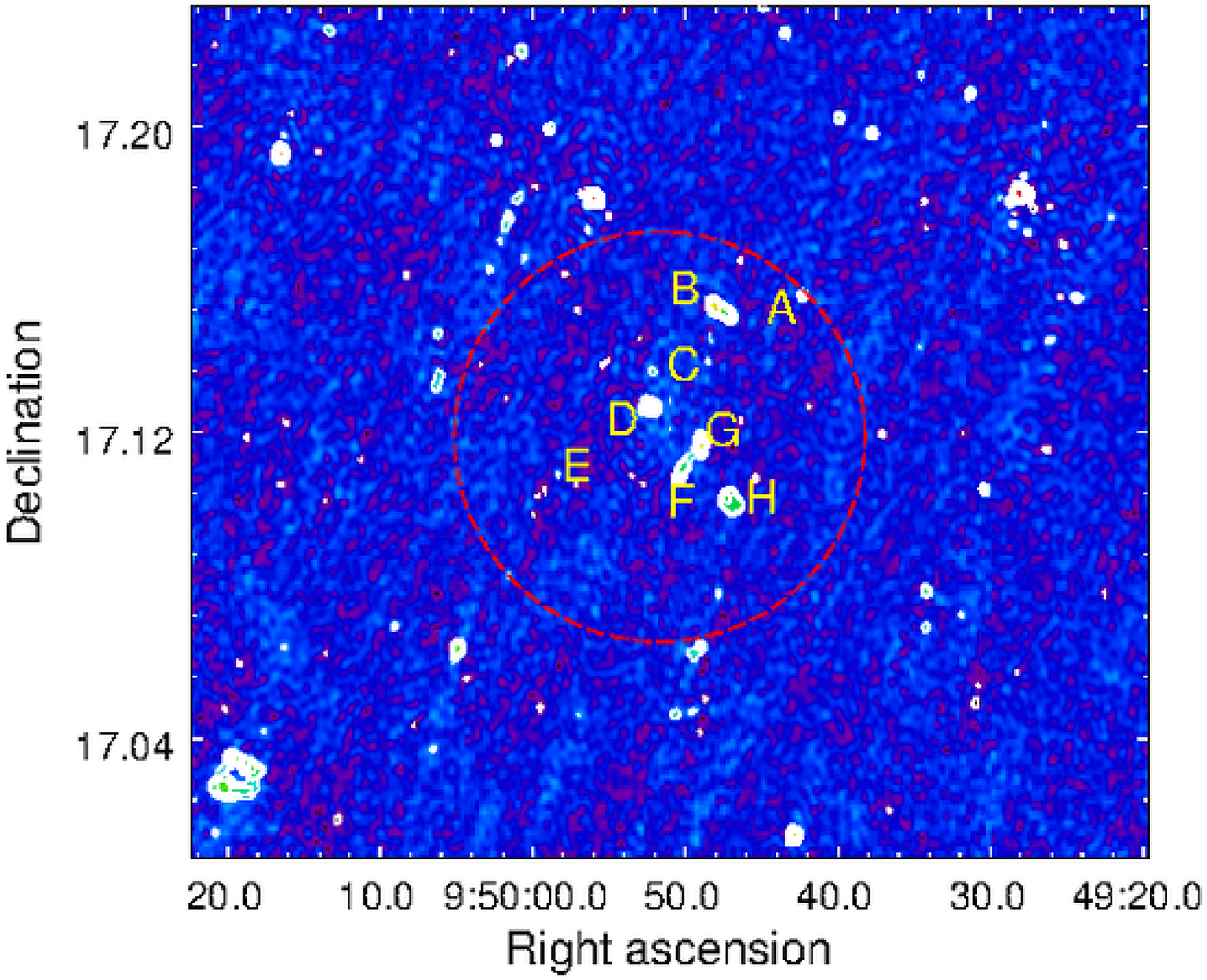}}
\put(50,0){\includegraphics[width=11cm]{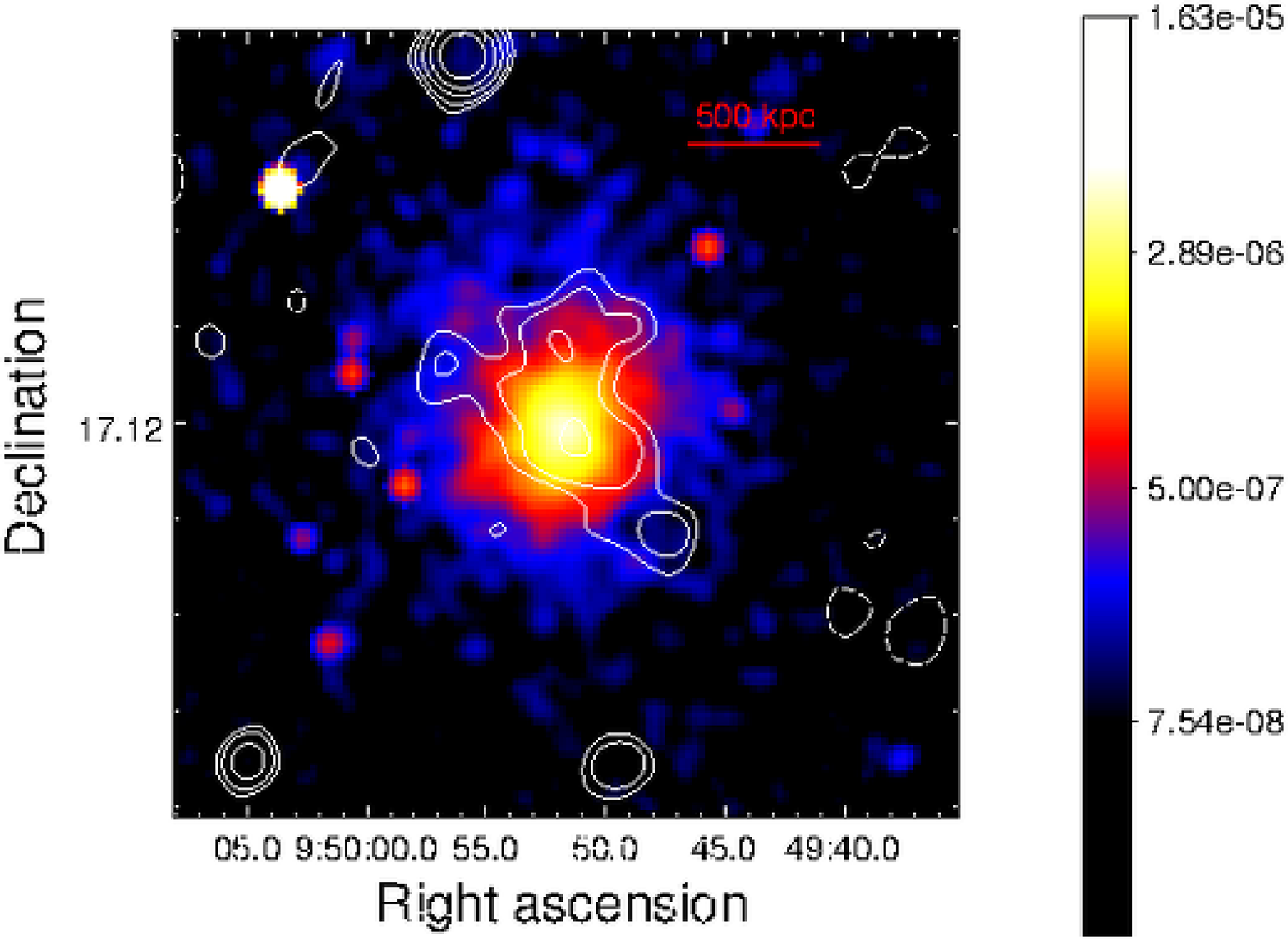}}
\end{picture}
\caption{RXCJ0949.9+1708: Left panel: HR image at 323 MHz in colours and iso-brightness contours. The beam is 9.8$" \times$7.9$"$. Contours start at 0.5 mJy/beam and are spaced by a factor 2. The first negative contour is dashed.The red dashed circle is centred on the X-ray centre of the cluster and has a radius of 1 Mpc. The sources subtracted in the LR image are marked by letters (A to H). Right panel: X-ray emission of the cluster from {\it Chandra} \citep{Ebeling10} in colours and iso-brightness contours from the radio LR image. The colorbar has units of photons cm$^{-2}$ s$^{-1}$ pixel$^{-1}$. The beam is 27.2$" \times$26.8$"$. Contours start at 0.7 mJy/beam and are spaced by a factor 2. The first negative contour is dashed.}
\label{fig:RXCJ0949}
\end{figure*}

\begin{figure}
{\includegraphics[width=8cm]{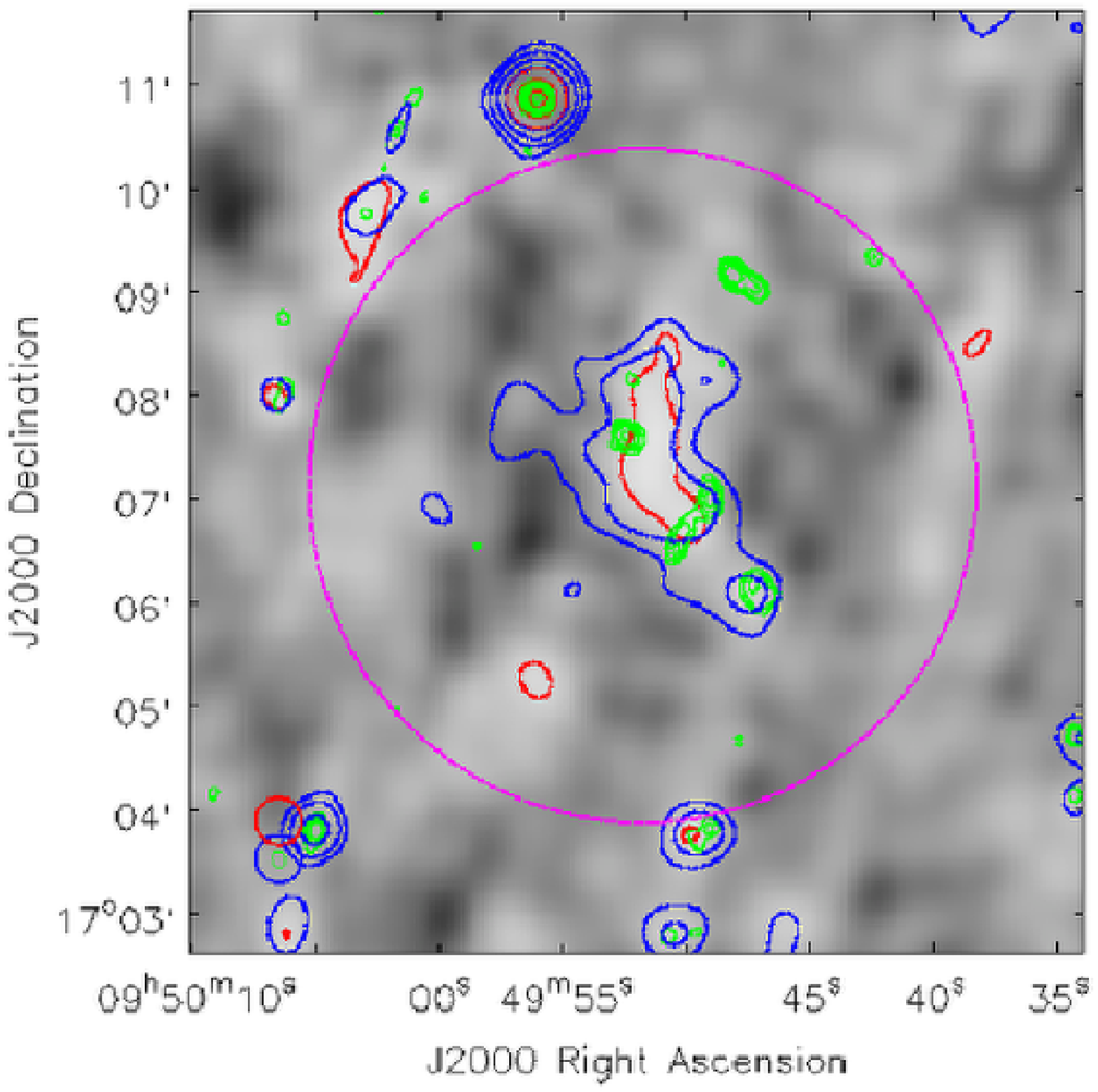}}
\caption{RXCJ0949+1708: Grey colour scale: LR image of the cluster, blue contours are like in Fig. \ref{fig:RXCJ0949}, right panel. Green contours are like in Fig. \ref{fig:RXCJ0949}, left panel.  Red contours are the residual of the HR image convolved with a beam as the one of the LR image. Red contours are drawn at $\pm 3\sigma$. The dashed-circle is centred on the cluster centre and has a radius of one Mpc. The restoring beams are shown in the bottom left corner.}
\label{fig:RXCJ0949_residual}
\end{figure}

\begin{figure}
\centering
\includegraphics [width=8cm]{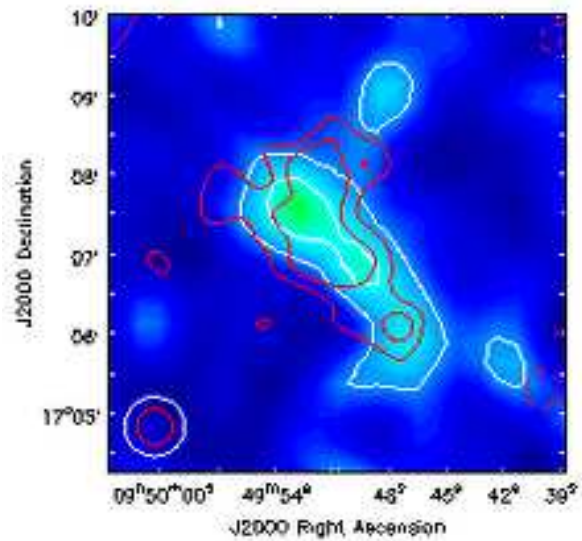}
\caption{RXCJ0949.9+1708: colours and white contours refer to the emission at 1.4 GHz from the NVSS. White contours are drawn at (-1,1,2,4) mJy/beam. The beam is $45" \times45"$. The first negative contour is dashed. Red contours refer to the LR GMRT image (as in Fig. \ref{fig:RXCJ0949}, right panel).}
\label{fig:RXCJ0949_nvss}
\end{figure}

\begin{figure*}
\vspace{100pt}
\begin{picture}(130,130)
\put(-210,-10){\includegraphics[width=11cm]{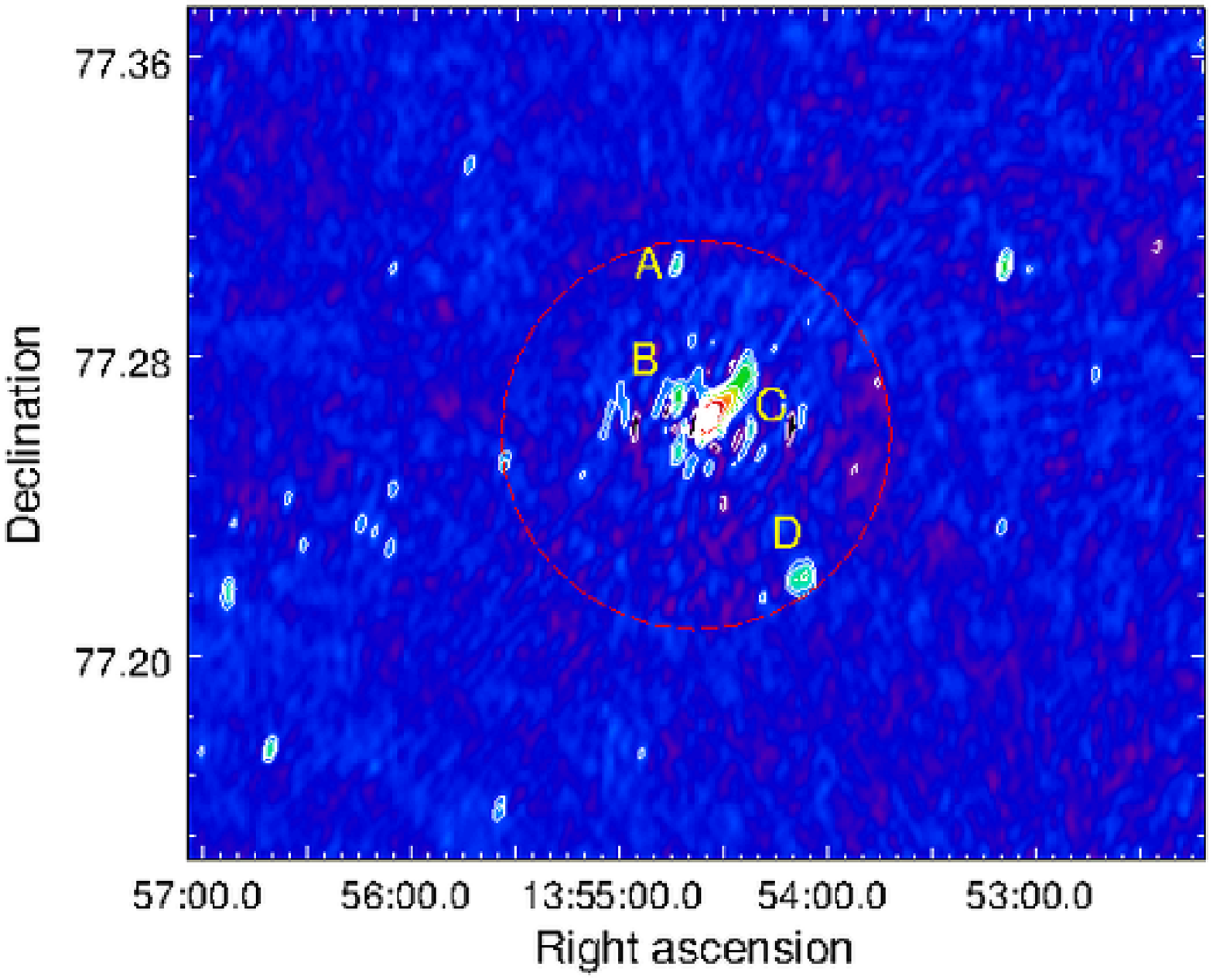}}
\put(45,0){\includegraphics[width=11cm]{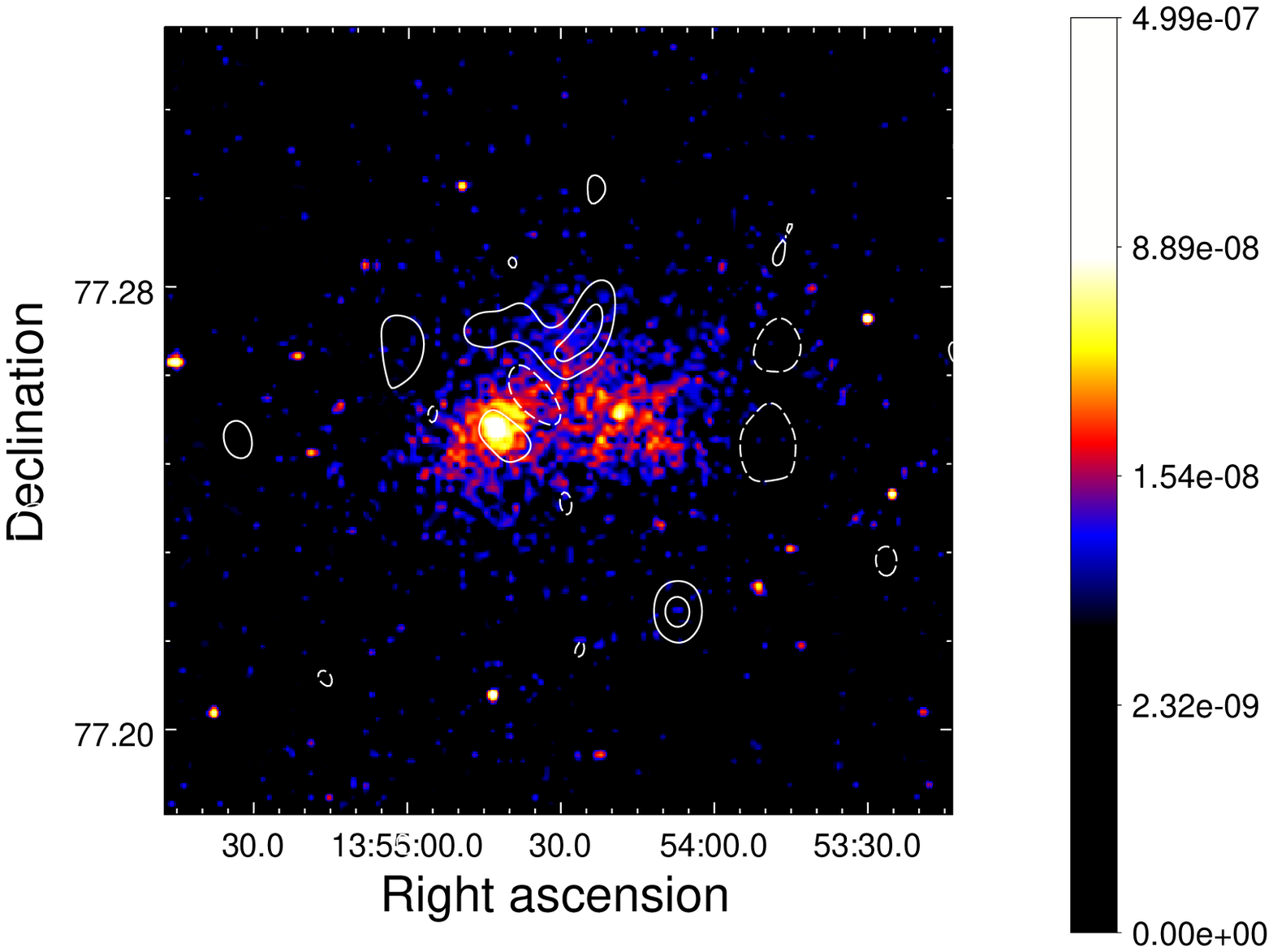}}
\end{picture}
\caption{RXCJ1354.6+7715. Left panel: HR image at 323 MHz in colours and iso-brightness contours. The beam is 19.4$" \times$878$"$. Contours start at 0.5 mJy/beam and are spaced by a factor 2. The first negative contour is dashed.The red dashed circle is centred on the X-ray centre of the cluster and has a radius of 1 Mpc. The sources subtracted in the LR image are marked by letters (A to D). Right panel: X-ray emission of the cluster from {\it Chandra} 0.5 - 2 keV energy band in colours and iso-brightness contours from the radio LR image. The colorbar has units of photons cm$^{-2}$ s$^{-1}$ pixel$^{-1}$. The beam is 33.7$" \times$25.6$"$. Contours start at 0.8 mJy/beam and are spaced by a factor 2. The first negative contour is dashed.  }
\label{fig:RXCJ1354}
\end{figure*}

\begin{figure}
\includegraphics[width=8cm]{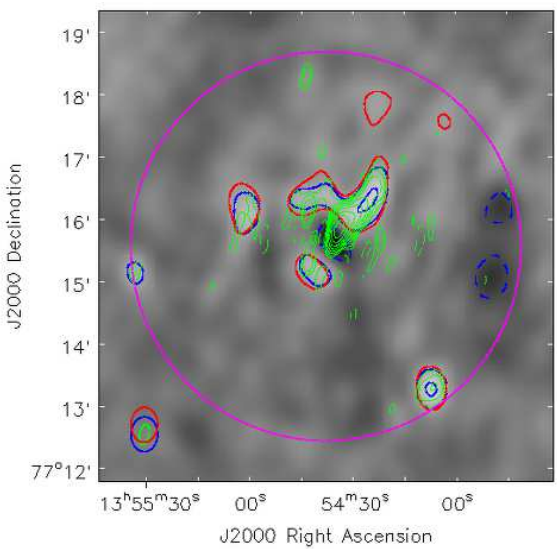}
\caption{RXCJ1354.6+7715: Grey colour scale: LR image of the cluster, blue contours are like in Fig. \ref{fig:CIZAJ1938}, right panel. Green contours are like in Fig. \ref{fig:CIZAJ1938}, left panel. Red contours are the residual of the HR image convolved with a beam as the one of the LR image. Red contours are drawn at $\pm 3\sigma$. The dashed-circle is centred on the cluster centre and has a radius of one Mpc. In the bottom left corner, the restoring beams are plotted. }
\label{fig:RXCJ1354_residual}
\end{figure}

\subsection{Abell 1443}
Little is known about Abell 1443 in the literature.  This cluster was one of the first detected by the Planck mission \citep{Planck11}.
\citet{Planck11} report a cluster mass, within $r_{500}$\footnote{$r_{500}$ is  defined as the radius at which the cluster density is 500 times the critical density.} 
of  $M_{500}=$7.7$^{+0.5}_{-0.6}$ $\times 10^{14} M_{\odot}$.  The ROSAT Brightest Cluster Sample 
reports an X-ray luminosity  $L_{[0.1-2.4 keV]} \sim 6.18 \times 10^{44}$ erg s$^{-1}$ \citep[][corrected for the cosmological model used in this paper]{rosat_BCS}.
{\it Chandra} archival observations are also available for this target. We have retrieved the {\it Chandra} observations from the archive, and processed them in the standard way (see Sec. \ref{discussion}).
 The X-ray image is shown in Figure \ref{fig:A1443}.
The X-ray morphology  appears elongated along  East-West, and no defined core is clearly visible. The gas distribution, as traced by the X-ray emission, is highly perturbed, indicating that the cluster is undergoing a merger event.

As shown in Fig. \ref{fig:A1443}, we detect peculiar diffuse emission in the GMRT 323 MHz image. 
The cluster hosts a $\Gamma$-shaped source located at $\sim$300 kpc from
the cluster X-ray centre, which does not have the morphology of a radiogalaxy (Figure \ref{fig:gammaSource}). 
Another patch of diffuse emission is located W of the $\Gamma$-shaped source (Figure  \ref{fig:gammaSource}), located at $\sim$700 kpc from the cluster X-ray centre.
After subtracting the sources A-G(Figure \ref{fig:A1443}, left panel), we have imaged the data at low-resolution, as described in Sec. \ref{sec:radio}.

Extended emission is detected, as shown in the low resolution image (Figure \ref{fig:A1443}, right panel).
We note that no residual emission from the sources A,B,C, and D is left in the low resolution image, indicating that the emission from the sources embedded in
the diffuse emission has been properly subtracted too.

The Largest Linear Size (LLS) of the emission is $\sim$ 1.3 Mpc. It departs  from the cluster centre toward W, encompassing the $\Gamma$-shaped source and the 
patch of diffuse emission, both detected in the high resolution image. 
Because of its peculiar morphology, a classification of this emission into the usual halo and relic categories is not straightforward.
The Western patch of diffuse emission could be a relic, as it is at the edge of the X-ray emission and it has a LLS of  $\sim$660 kpc. 
Given its morphology and  the fact that no component is visible in the WISE 3.4 image (Fig. \ref{fig:A1443}, \citealt{WISE}) the possibility that we are seeing a radio galaxy is less convincing.

The emission located centrally could be instead classified as a radio halo, as it has a LLS of $\sim$1.1 Mpc. 
However, the emission itself is peculiar because of its morphology and it contains a bright $\Gamma$-shaped source.
In another case (MACSJ0717+3745, \citealt{Bonafede09b} \citealt{vanweeren09}) a brighter part, possibly a relic or a bright filament,  
has been found embedded into the radio halo emission.
Spectral index and polarisation information are needed to discriminate the origin of the $\Gamma$-shaped source and understand the nature
of the whole diffuse emission.
Details about the radio properties of the emission detected in this cluster are listed in Table \ref{tab:sources}.

To verify that the diffuse emission is not affected by the incomplete subtraction of the sources  detected at high-resolution (A-G in Fig. \ref{fig:A1443})
we have convolved the residual of the high-resolution image with a Gaussian having the major and minor axis as the restoring beam of the LR image (Fig. \ref{fig:A1443_residual}).
 In this image, the HR contours are plotted over the LR ones, and over the contours of the residuals of the HR image convolved at low resolution.
From this image we conclude that the sources A,B,C,D and F are properly subtracted. Some residual is detected corresponding to the position of source G, on top of the relic.
This emission contributes with 0.2 mJy to the relic emission. Therefore, it does not affect our estimate of the relic flux density.

\subsection{CIZAJ1938.3+5409}
\label{sec:ciza}
The cluster CIZAJ1938.3+5409 is located in the so-called {\it ``zone of avoidance"}, and was discovered by \citet{CIZA}
by inspecting X-ray data from the RASS Bright Source Catalog \citep{rosat_BCS}.
The authors report an X-ray luminosity of $L_{[0.1-2.4 keV]} \sim 10.89  \times 10^{44}$ erg s$^{-1}$
(corrected for the cosmological model adopted in this work).
The cluster mass, as reported by Planck \citep{Planck13} is $M_{500} = 7.5^{+0.4}_{-0.3} \times 10^{14} M_{\odot}$.
Therefore, it is quite a massive cluster, but unfortunately nothing is known in the literature about its dynamical state.

{\it Chandra} observations are available for this target (Jones et al., 2015, in preparation). 
The X-ray and  radio emission of the cluster are shown in Fig. \ref{fig:CIZAJ1938}. 

\begin{figure}
\centering
\includegraphics[width=8.2cm]{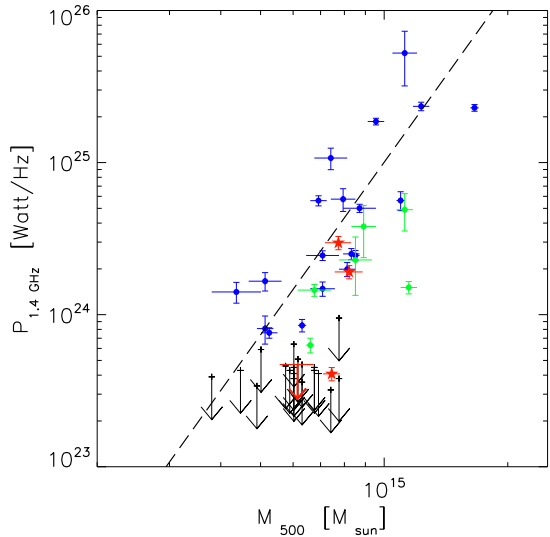}
\caption{Correlation between the radio power of radio halos at 1.4 GHz and the cluster mass $M_{500}$  as derived by SZ measurements \citep{Planck13}.  Blue points are the radio halos used to derive the best-fit line (shaded line). Green points are halos with spectral index $\alpha>1.5$. Arrows are upper limits (points, arrows and bet-fit lines taken by see Cassano et al. 2013). Red filled stars are the radio halos presented in this paper, for which a spectral index $\alpha=1.2$ is assumed. The red arrow refers to the upper limit on RXCJ1354.6+7715 (halo size $\sim$ 700 kpc, 3$sigma_{LR}$ detection threshold, see text for details). }
\label{fig:corr}
\end{figure}

After the source subtraction, we discover extended radio emission, located at the cluster centre and
having a size of $\sim$ 720 kpc (Fig. \ref{fig:CIZAJ1938}, left panel). 
The flux density of the emission is 11 mJy, which translates into a 
power of $2.4\pm 0.3 \times 10^{24}$ W Hz$^{-1}$ at 323 MHz.
The size and the position of the emission would classify it as a radio halo. We note that the corresponding power at 1.4 GHz would be
$P_{1.4 \rm{GHz}} \sim 4\times 10^{23} \rm{~W~Hz^{-1}}$, which is unusually low for such a massive cluster\footnote{we assume $\alpha=1.2$}. 
Indeed, the halo is  at least 8-10 times under-luminous in radio for its SZ signal and X-ray power, respectively (see Fig. \ref{fig:corr}). 
The radio power at 1.4 GHz would place the halo in the ``upper-limit" region
of the $P_{1.4\, {\rm GHz}}-M_{500}$ plane.

To verify that the diffuse emission is not due to the the incomplete subtraction of the sources  detected at high-resolution (A-H in Fig. \ref{fig:CIZAJ1938})
we have convolved the residual of the high-resolution image with a Gaussian having the major and minor axis as the restoring beam of the LR image (Fig. \ref{fig:CIZAJ1938_residual}).
 In this image, the HR contours are plotted over the LR ones, and over the contours of the residuals of the HR image convolved at low resolution.
From this image we conclude that the sources A,B,C,D,F,G and H are properly subtracted, as no residual emission is detected neither in the LR image nor in the residual image. 
Some residual emission is detected corresponding to source E. This emission accounts for 0.2 mJy and it is only confined to the source position, hence it cannot explain the 11 mJy emission on larger scale, detected in the LR image.

\begin{figure}
\centering
\includegraphics[width=8cm]{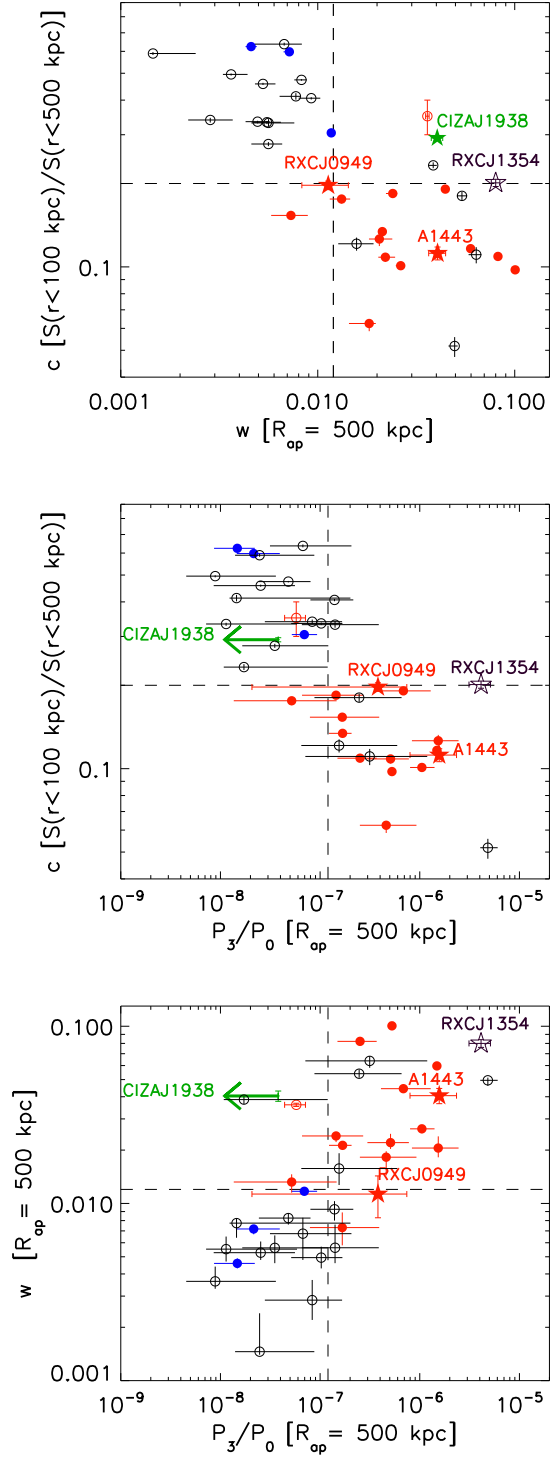}
\caption{The morphological diagrams $c-w$ (top), $c-P_3/P_0$ (middle) and $w-P-3/P_0$,  adapted from Cassano et al. (2010). Red  filled symbols refer to radio halos (red filled circles are from Cassano et al. (2010), red empty circle is CL1821+643 from Bonafede et al (2014), red filled stars are the radio halos discovered in this paper. Blue filled dots refer to to mini halos, black circles are clusters with no radio emission (from Cassano et al. 2010). The green star and arrows refer to the low power radio halo in CIZAJ1938.3+5409.  Arrows are upper limits to the $P_3/P_0$ value.  }
\label{fig:mom}
\end{figure}

\subsection{RXCJ 0949.8+1708}
The cluster RXCJ 0949.8+1708, also known as MACS J0949.8+1708,  is an X-ray luminous cluster, discovered in  X-rays
by the ROSAT satellite \citep{rosat_BCS}.  The {\it Planck} satellite detected the cluster through the SZ effect \citep{Planck11}, and found it to be massive, with 
$M_{500} =  8.2 \pm0.6  \times10^{14} M_{\odot}$ \citep{Planck13}.

\citet{Ebeling10} have analysed the cluster in detail using  the UH 2.2m optical telescope and X-ray {\it Chandra} observations. They derive an X-ray luminosity $L_{x [0.1 - 2.4 keV]}=10.6 \pm 0.6 \times 10^{44} \rm{erg \, s^{-1}}$ 
within $R_{500}$, and a temperature $T=8.9 \pm 1.8$ keV.  

In Fig. \ref{fig:RXCJ0949} the X-ray emission of the cluster is shown. 
The MACS clusters in \citet{Ebeling10} have been classified with a morphological code from 1 o 4,  1, meaning pronounced cool-core and good optical-X ray alignment, and  4  meaning  multiple X-ray peaks and no cD galaxy. 
RXCJ 0949.8+1708 is given a morphological code 2, meaning that although no pronounced cool-core  is present, the X-ray emission shows a good optical-X-ray alignment  and concentric X-ray contours.

This cluster has already been observed by \citet{Venturi08} at 610 MHz (the author refer to the cluster with the name Z2261), who found positive residuals after the subtraction of the sources and indicated the cluster as a candidate radio halo.

In Fig. \ref{fig:RXCJ0949} the radio image from our new GMRT observations is shown. 
We detect diffuse radio emission located at the cluster centre and with a LLS of $\sim$1 Mpc, confirming that the positive residuals detected by \citet{Venturi08} are 
part of more extended radio emission.
We classify the emission as a radio halo. It is elongated in the SW-NE direction, and it does not follow the emission of the gas. 

To verify that the diffuse emission is not affected by the incomplete subtraction of the sources  detected at high-resolution (A-H in Fig. \ref{fig:RXCJ0949})
we have convolved the residual of the high-resolution image with a Gaussian having the major and minor axis as the restoring beam of the LR image (Fig. \ref{fig:RXCJ0949_residual}).
From this image we conclude that the sources are properly subtracted. Some residual is detected on top of the brightest region of the halo. Likely, it is the brightest part of the radio halo, which is
on a scale smaller than 1 Mpc and, as such, is not filtered-out in the HR image. The flux density corresponding to this emission is $\sim$ 0.5 mJy, hence, even if it came from the individual radio sources
it would not affect the estimate of the halo flux density (see Tab. \ref{tab:sources}).

\subsection{RXCJ1354.6+7715}
RXCJ1354.6+7715, also known as MACS J1354.6+7715, is an X-ray luminous galaxy cluster discovered by \citet{NORAS}. 
They report an X-ray luminosity in the energy band 0.1-2.4 keV of $L_{x [0.1 - 2.4 keV]}= 9.4 \times 10^{44} \rm{erg \, s^{-1}}$ (corrected for the cosmological model used in this work). 

The {\it Planck} satellite detected the cluster through the SZ effect \citep{Planck11}. \citet{Planck13} report $M_{500} = 6.2 \pm0.6  \times 10^{14} M_{\odot}$.

\citet{2010MNRAS.406.1318H} have analysed the matter substructure in the cluster using strongly lensed arcs detected through HST observations.
Their analysis suggests the existence of two separate galaxy concentrations. They conclude that this cluster could be during some stage of a merger, with a considerable amount of substructure.

{\it Chandra} observations are available in the archive. We have calibrated them in the standard way (see Sec. \ref{discussion}) and in Fig. \ref{fig:RXCJ1354} the
X-ray emission is shown. In agreement with the optical analysis by \citet{2010MNRAS.406.1318H}, two gas concentrations are clearly detected.

In Fig. \ref{fig:RXCJ1354} the radio emission from the cluster at 323 MHz is shown. A bright radio galaxy (C in Fig. \ref{fig:RXCJ1354}, left panel) 
is located at the centre of the cluster. Its total flux is $\sim$159 mJy. It has a nucleus and a lower surface brightness tail. Residual calibration errors are present around the radiogalaxy. The  presence of the tail and the residual calibration errors  make the subtraction of the clean components of the radiogalaxy C difficult.

 Some residual emission is present in the low-resolution radio image, which is likely due to the convolution at lower resolution
of the residual calibration errors of the radio galaxy C and to residual emission from the tail. In order to confirm this, we have convolved the residuals, obtained after the cleaning of the high-resolution image, with a Gaussian having the major and minor axis as the restoring beam of the low-resolution image. The comparison is shown in  Fig.  \ref{fig:RXCJ1354_residual}.  Hence, despite the massive and likely merging cluster, no radio emission in form of radio halo or relic is present at the  sensitivity level reached by our observations.

It must be noted that RXCJ1354.6+7715 is the least massive clusters among those analysed in this paper. Putting an upper limits to the possible radio emission below our detection threshold is not trivial, because of the 
presence residual calibration errors connected with the radio galaxy C, and more generally  because of the non-universal surface brightness distribution of radio halos.
However, considering a conservative detection threshold of 3$\sigma_{LR}$ over a circle of 350 kpc  radius, we can exclude the presence of  radio emission  on a scale$\geq$ 700 kpc with a flux density $S \sim 5$ mJy, which translates into an upper limit of $P_{1.4 \, {\rm GHz}} < 5 \times 10^{23}$. This value is a factor $\sim$5 below the $P_{1.4 \, {\rm GHz}}$-SZ correlation.

\section{Discussion}
\label{discussion}
\subsection{Substructure analysis}
\label{substructure}
All the clusters analysed in this work have {\it Chandra} archival observations. 
X-ray images are a powerful tool to understand the cluster dynamical state, complementary to optical observations.

 We have reprocessed the {\it Chandra} observation of the clusters  using the latest calibration tables (CALDB 4.5.9) in CIAO. 
Following \citet{Cassano10}, the event files have been binned by a factor four in order to undersample the {\it Chandra} PSF. 
We have selected the 0.5-2 keV energy band, and  we have verified that more than 2000 counts were detected inside the region of interest (a circle of 500 kpc radius) to allow for a morphological analysis.
The images have been normalised by the exposure map, and the sources in the cluster field have been visually identified and excluded from the analysis.

In order to quantify the dynamical status of the clusters,  we compute three dynamical indicators, extensively used in the literature \citep[e.g.][]{Boehringer10}:  the power ratio $P_3/P_0$, the centroid shift, $w$, and the concentration parameter, $c$.

\citet{Cassano10} have used these methods, finding that clusters with and without radio halos occupy different regions in the morphological diagrams. Recently, we have found a first outlier \citep{Bonafede14b} in the case of CL1821+643.

The morphological indicators $P_3/P_0$, $w$, and $c$ have been computed within an aperture radius $R_{ap}=500$ kpc, and considering as centre the centroid of the X-ray emission ($P_3/P_0$ computation) and the cluster X-ray peak ($c$ and $w$ computation), to have a fair comparison with the clusters analysed by \citet{Cassano10}. We note that the clusters presented in this work have mass and redshift comparable to the clusters analysed by \citet{Cassano10}, hence the comparison makes sense.
The errors and the photon bias have been estimated through Monte-Carlo simulations, following  the approach of \citet{Boehringer10}. 

The centroid shift  method \citep{Maughan08} measures the standard deviation of the projected separation between the X-ray peak and the centroid in units of $R_{ap}$, 
computed within circles of increasing radius (from 0.05$\times R_{ap}$ to $R_{ap}$, in steps of 0.05$\times R_{ap}$). 
The centroid shift, $w$, has been derived computing the X-ray weighted centroid  within circles of increasing radius. 

The concentration parameter, $c$, measures the ratio between the X-ray surface brightness within 100 kpc over the surface brightness within 500 kpc \citep{Santos08}. It is higher
for cool core and regular clusters, while it is lower for merging clusters.

The power ratio is a multipole decomposition of the two-dimensional projected mass distribution \citep{1995ApJ...452..522B}.  
We have determined the power ratio $P_3/P_0$, which is the lowest power ratio moment providing a clear substructure measure. Following \citet{Boehringer10}, momenta are computed excising the inner 0.05$\times R_{ap}$. In the computation of $P_3/P_0$, the photon bias takes into account the spurious amount of substructure that could contribute to the power ratio. As noted by \citet{Boehringer10}, also very regular clusters would show some substructure because of the photon noise. Since we are comparing here with the analysis by \citet{Cassano10}, the bias will not be taken into account in the following analysis. 
However, the photon bias gives a larger contribution to regular clusters than to perturbed ones.
Hence, once it is taken into account in the power ratio analysis, the separation of merging and regular clusters along the $P_3/P_0$ axis would likely be more evident.

In Fig. \ref{fig:mom} we show the position of the clusters in the $P_3/P_0-c$, and $w-c$ diagrams, taken from \citet{Cassano10}, \citet{Bonafede14b}. 

\subsubsection{Abell 1443}
The substructure analysis confirms the visual impression from the X-ray image. 
We obtain $w=0.0406 \pm 0.004$, $c=0.112 \pm 0.006$, and $P_3/P_0=( 1.6 \pm 0.7) \times 10^{-6}$.
The dynamical status of the cluster is highly perturbed, and the cluster falls in the
merging quadrant of the $P3-P0$, $w$, $c$ diagrams\footnote{As expected, in this case the bias would not change the value of $P_3/P_0$ dramatically. Taking the photon bias into account one would obtain $P_3/P_0=( 1.5 \pm 0.7) \times 10^{-6}$ }.  As expected for such massive and merging clusters, 
diffuse radio emission has been discovered.

\subsubsection{CIZAJ1938.3+5409}
We obtain an upper limit for the value of $P_3/P_0 < 3.8 \times 10^{-8}$, and a rather high value of $c=0.292 \pm 0.006$ that are typical for 
non-merging clusters. Instead, the shift parameter $w=0.040 \pm0.003$ is typical of merging clusters.
Although a major merger can be reasonably excluded, the analysis is not conclusive of the cluster dynamical status.

In Fig. \ref{fig:mom} the cluster position in the morphological planes is shown. We note the peculiar position of the cluster in the 
diagrams. While in the  plane $c-P_3/P_0$ the clusters sits among the non-merger ones, it lies in an almost empty quadrant in the $c-w$ and $w-P_3/P_0$ planes.

We note that the cluster morphological indicators are very similar to the one of CL1821+643 \citep{Bonafede14b}. However, the radio emission in the two clusters 
are rather different. CL1821+643 hosts a radio halo with a power that follows the $P_{1.4\, {\rm GHz}}-L_X$ and $P_{1.4\, {\rm GHz}}-\rm{SZ}$ correlations, while
the radio halo discovered in CIZAJ1938.3+5409 is at least 8 times below the  $P_{1.4\, {\rm GHz}}-\rm{SZ}$   correlations.
 
Recently, \citet{2015arXiv150500778K}  has shown that Planck SZ measurements might be contaminated by CO Galactic emission. This would result in an over estimate of the
cluster mass from SZ measurements, that can be relevant for clusters close to the Galactic plane. In the case of CIZAJ1938.3+5409, 
a mass  $M_{500} \sim 4 \times 10^{14} M_{\odot} $ would bring the radio halo on the $P_{1.4\, {\rm GHz}}-\rm{SZ}$ correlation. As the error on Planck masses are not quantified yet, a more precise estimate of the possible error in this cluster cannot be derived. However, we note that the cluster does not fall in the most CO contaminated regions derived by  \citet{2015arXiv150500778K}. This suggests that the mass derived by Planck is a good estimate of the cluster mass.

{Since the cluster total Luminosity in the band 0.1-2.4 keV is known (\citealt{CIZA}, see Sec. \ref{sec:ciza}), we have checked the position of the radio halo in the $P_{1.4 \, GHz} - L_X$ plane \citep{Kale15}. Also in this diagram the radio halo is a factor $\sim$ 10 below the $P_{1.4 \, GHz} - L_X$ correlation, and falls in the upper-limit region. This provides an independent evidence - with respect to the Planck mass measurement - that the cluster hosts a radio halo of low power for its mass.}

 In the next Section, we discuss the possible origin for the low power radio halo in CIZAJ1938.3+5409.

\subsubsection{RXCJ0949.9+1708}

For RXCJ0949.8+1708, the morphological analysis is not conclusive of the cluster dynamical status. 
We obtain $w=0.0113 \pm 0.003$, $c=0.197 \pm 0.007$, and $P_3/P_0=( 3.8 \pm 3.6 ) \times 10^{-7}$.
As can be seen from Fig. \ref{fig:mom}, $w$, and $c$ are at the edge of the quadrant among radio quiet and radio loud clusters, and so does also $P_3/P_0$,
within the error. We note that if the bias was taken into account in the $P_3/P_0$ computation, we would only be able to put an upper limit $P_3/P_0<3.5 \times 10^{-7}$.

The cluster sits in the central regions of the morphological diagrams, in between regular and merging clusters.
The values of $w$, $c$,  $P_3/P_0$, and the presence of a BGC at the cluster centre, suggest that no major merger is occurring.
However, present data do to allow us to exclude a major merger along the line of sight. Additional optical or weak lensing
data are required to assess the cluster dynamical status.
 
{ We have estimated the spectral index of the halo in RXCJ0949.8+1708 using NVSS and VLSSr  (VLA Low-Frequency Sky Survey Redux Source Catalog, \citealt{VLSSr}) images. In Fig. \ref{fig:RXCJ0949_nvss} the radio emission from NVSS is shown, superimposed onto the GMRT LR image.
We have estimated the spectral index in the region where both images have signal above three times the noise. Since it is not possible to estimate the contribution of the sources (A to H in Fig. \ref{fig:RXCJ0949}) in the NVSS image, only an upper limit to the halo flux can be derived. Since no emission is present in the VLSSr image above three times the noise we have put an upper limit to the halo flux. Assuming that the spectrum is described by a single power-law between 74 MHz and 1.4 GHz, we can constrain the spectral index in the range $0.6 < \alpha < 1.5$. Hence, we can only exclude a very steep spectrum radio halo.}

\subsubsection{RXCJ1354.6+7715}
The results from the substructure analysis are $P_3/P_0=(4 \pm 1) \times 10^{-6}$, $w=0.080 \pm 0.004$, and 
$c=0.201 \pm 0.006$, which place the cluster among the merging ones, although the high value of $c$ indicates that the core has not been disrupted by the merger. Alternatively, the high value of $c$
could indicate a second merger along the line of sight.

These results are in agreement with the optical analysis by \citet{2010MNRAS.406.1318H}.\\
No radio halo is detected at the sensitivity threshold of our observations, and under some assumptions (see Sec. \ref{sec:clusters} ) we cannot exclude the
presence of diffuse radio emission a factor five below the  $P_{1.4\, {\rm GHz}}-\rm{SZ}$ correlation.

\subsection{CIZAJ1938.3+5409: the first hadronic halo?}

The radio halo  discovered in CIZAJ1938.3+5409 is the first direct detection of radio emission in a radio off-state cluster. 
As it is at least eight times under-luminous in radio for its SZ signal, we discuss two possibilities for the origin of the radio emission:
\begin{enumerate}

\item{ {  The observed emission in  CIZAJ1938.3+5409 could be due to secondary electrons originating from hadronic interactions between thermal and non-thermal protons \citep{2004A&A...426..777P,Ensslin11}. We note that the mentioned works aimed at explaining radio halos with a higher power than the one observed in CIZAJ1938.3+5409.
However, they can be easily extrapolated at lower radio powers because the particle spectrum predicted by the models is a simple power-law. The extrapolation at lower radio power has been alredy done e.g. by \citet{BL11}.  According to this work, the electrons originating from hadronic collisions could produce radio emission that is a factor $\sim$ ten  below the $L_X -P_{1.4\, {\rm GHz}}$ correlation.}
 {\it Fermi} upper limits indicate that the ratio of non-thermal CR protons to thermal energy densities is lower than few $\%$ \citep{Ackermann14}.\\
\indent Stacking a sample of clusters in radio,  \citet{Brown11}  have claimed the first detection of the radio-off state of clusters.
In order to better understand the possibility that this halo is produced by secondary electrons, we have investigated a set  of simple models for CIZAJ1938.3+5409. 
As a first approximation, we have assumed a $\beta$-model for the gas density, rescaled to the mass of the cluster. We have assumed an isothermal temperature of $T \approx 6.4$ keV as indicated by X-ray data,  and we have computed the profile of cosmic ray (CR) protons by requiring that the energy in CR is a fixed fraction of the gas energy, $\epsilon_{\rm CR}$. \\
\indent Furthermore,  we have modelled the magnetic field of the cluster as we have done in \citet{Bonafede13}. Specifically, we have generated a power-law spectrum distribution of vector potential in the Fourier space for a $200^3$ grid, randomly drawn from the Rayleigh distribution, and we have computed the magnetic field in the real space as $\bf B=\nabla \times \bf A$., ensuring $\nabla \cdot \bf B=0$ by construction. 
We have assumed that the maximum coherence scale of the magnetic field is 34 kpc, that the power-law of fluctuations is a Kolomogorv spectrum and that the  average magnetic field strength in the central $\rm Mpc^3$ is $\langle B \rangle =2 \, \mu$G, as found for Coma \citep{Bonafede10}.\\

\indent  Since the radio spectrum of the halo is not well constrained, we explore a few possibilities for the spectrum of the cosmic ray protons, and adjusted the energy budget of CRs so that the secondary radio emission matches our observation.  We compute the hadronic $\gamma$-ray emission, following \citet[][]{pe04,donn10}, with the only difference that for the hadronic cross-section we use the parametrisation of the proton-proton cross section given by \citet{2006PhRvD..74c4018K}, see also \citet{Vazza15_p} for details. The secondary emission is based on \citet{DolagEnsslin2000}. 

Results are listed in Table \ref{tab:tab_gamma}.
All models with a spectrum of CRs $p \leq -3$ (with $dN/dE \propto E^{p}$) produce hadronic emission below the deepest limits by {\it Fermi}, which
are at the level of $\sim 10^{-10} \rm ph/(cm^2 s)$ (e.g. Huber et al. 2013, Ackermann et al. 2014). The model with $p = -3.1$  produces emission above the detectability of {\it Fermi}. However, the energy budget of CRs nearly equals the thermal energy, making the model unrealistic.\\
\indent A direct observation of this cluster with {\it Fermi} is unpractical due to its location in the Zone of Avoidance. Our calculations suggest that a hadronic origin for the radio halo in CIZAJ1938.3+5409 is consistent with the current {\it Fermi} limits.\\
\indent However, we note that if the cluster was at the luminosity distance of Coma ($\sim$ 99 Mpc) it should be detected in the $\gamma$-rays only if $p \leq -2.3$, i.e.   $\alpha \geq 1.2$.
These estimates must be taken as approximative because of the uncertainties in both the magnetic field distribution and strength, which translate into an uncertainty in the normalisation of the energy budget of CRs for a given spectrum. 
For instance,  for a magnetic field $\langle B \rangle =0.1 \, \mu$G,  {\it Fermi} upper-limits would be violated for $p=-2.1$.\\

Our calculations suggest that if the halo has a hadronic origin, some clusters of the same mass as CIZAJ1938.3+5409 in the nearby Universe might host radio and $\gamma$-ray emission just above the current level of detectability.  These halos do not represent the classic giant radio halo population, but rather a much broader population of ``radio-quiet" objects, that still have to be detected.
 }

\item{ The second possibility to explain the radio emission in CIZAJ1938.3+5409 is that we are observing a radio halo in its rapid transition phase between the upper-limits region and the $P_{1.4\, {\rm GHz}}$-SZ correlation.
A direct test, as done for the hadronic origin of the radio emission above, is not possible here  because of the many
free parameters. However,  we can make some statistical considerations in order to establish if the low power radio halo could be due to the same mechanism that powers radio halos on the $P_{1.4\, {\rm GHz}}$-SZ correlation.

In the framework of re-acceleration models, merger-induced turbulence lights up the radio emission, lifting the clusters from the radio off-state region
of the $P_{1.4\, {\rm GHz}}$-SZ plane up to the radio on-state, on the $P_{1.4\, {\rm GHz}}$-SZ correlation. The steepest spectrum radio halos are expected to be below the correlation, as they
represent the latest stage of the radio halo life \citep{Donnert13}.  The halo in CIZAJ1938.3+5409 could be in the very early or very late stage of its radio-on phase.
In both cases, its spectrum should be  steep ($\alpha \geq 1.5$).
As no emission is detected in the VLSSr   image, we can only put an upper limit to the spectral index $\alpha < 1.9$. We cannot exclude that the radio halo has a very steep spectrum and that we are observing it just before it enters its radio-off state. 

According to the simulations by \citet{Donnert13}, clusters should spend $\sim$ 0.1- 0.2 Gyr in the region of the $P_{1.4\, {\rm GHz}}-L_X$ plane between the upper limits and the radio power measured for CIZAJ1938.3+5409.
Clusters as massive as   CIZAJ1938.3+5409 formed $\approx$ 6 Gyr ago \citep{Giocoli07}. 
Hence, assuming that all the clusters follow the same path in the $P_{1.4\, {\rm GHz}}-L_X$ plane, one would expect to find at maximum one cluster in $\approx$ 20-40 objects of the same mass and at the same redshift in the transition region, which is not at odds with current statistics. A proper estimate should account for the different merging rate at different redshifts. Hence, the number we find must be regarded as an upper limit.
 
 If the radio emission detected in CIZAJ1938.3+5409 is due to the same mechanism that powers the halo on the $P_{1.4\, {\rm GHz}}$-SZ correlation, the cluster should be undergoing a merger, which has not left a strong imprint in the X-ray emission (see Sec. \ref{substructure}). 
We note that the cluster lies in a rather empty quadrant of the morphological diagram $c-w$ (Fig \ref{fig:mom}), in agreement with the hypothesis that we are observing it during a short phase, in transition between merging and non-merging clusters.}

The two possible scenarios described above make very different predictions: if we are observing a very rare object in its rapid transition between the radio quiet and radio-loud phase, we expect to observe few objects of this type  in large cluster samples. Instead, if the halo has a hadronic origin, radio emission will be detected in every massive cluster (the upper-limits in Fig. \ref{fig:corr}).
The detection of hadronic halos would also allow to constrain the energy budget in CR proton in the ICM.

Future deep radio surveys performed e.g. with LOFAR \citep{LOFAR}  and later on with the Square Kilometer Array will shed light 
on this \citep{Cassano15}.

\end{enumerate}

\begin{table*}
\caption{Parameters of our models of hadronic secondary emission for PSZ1 G086.47+15.31, always assuming a Coma-like magnetic field configuration, with $\langle B \rangle=2 \mu G $ in the central Mpc$^3$. See text for details.} 
\centering \tabcolsep 2pt 
\begin{tabular}{c|c|c|c|c|c|c}
model & $p$ & $\alpha$& $\epsilon_{\rm CR}$ & $P_{\rm 325 ~ MHz}$ & $P_{\rm 0.5-100 ~ GeV}$ & $F_{\rm 0.5-100 ~GeV}$ \\
         &                         &                 &                  &     $[W/Hz]$   &   $[erg/s]$     &    $[ph/(cm^2 s)]$ \\ 
         \hline
 1        &        -3.1          &   1.7 & 0.91      &   $2.4 \times 10^{24}$  & $5.2 \times 10^{42}$ & $3.9 \times 10^{-10}$\\  
  2       &        -2.9          &   1.6 & 0.13      &   $2.4 \times 10^{24}$  & $8.5 \times 10^{41}$ & $6.5 \times 10^{-11}$\\
   3      &        -2.7         &   1.5 & 0.018       &   $2.4 \times 10^{24}$  & $1.4 \times 10^{41}$ & $1.1 \times 10^{-11}$\\
    4     &        -2.5          &  1.3 &  0.0093      &   $2.4 \times 10^{24}$  & $2.3 \times 10^{40}$ & $ 1.7 \times 10^{-12}$\\
     5    &        -2.3         &  1.2 &  0.0004       &   $2.4 \times 10^{24}$  & $3.6 \times 10^{39}$ & $2.7 \times 10^{-13}$\\
      6   &        -2.1          & 1.1 & 0.0001           &  $2.4 \times 10^{24}$  & $5.5 \times 10^{38}$ & $4.2 \times 10^{-14}$\\  
\hline
\hline
\end{tabular}
\label{tab:tab_gamma}
\end{table*}

\section{Conclusions}

\label{sec:conclusions}
In this work we have presented and analysed the radio emission from four galaxy clusters selected from the Planck catalog \citep{Planck11}.
The clusters are all massive, with $M_{500} > 6 \times 10^{14} M_{\odot}$ and are good candidates to host 
radio diffuse emission. We have used the same procedure to reduce and analyse the radio emission from the clusters, and with the help of {\it Chandra} archival data we have performed a substructure analysis to quantify the dynamical status of the clusters. We have detected three radio halos and we have placed one new upper limit in the $P_{1.4\, {\rm GHz}}-SZ$ plane.

Detection of diffuse radio sources in three out of four newly observed clusters, or in four out of five if we consider the previously published results on CL1821$+$643 (Bonafede et al. 2014) selected from the same SZ sample \citep{Planck11}, is indicative of a high fraction of occurrence of radio halos in SZ selected samples. Such high percentage have previously been suggested based on the analysis of NVSS radio images of Planck clusters \citep{SommerBasu}. The small sample size of the current work does not permit to make  any statistical predictions. 
However, the first Planck catalog \citep{Planck11} shows no net bias with respect to the later Planck SZ catalogs \citep[e.g.][]{Planck13}] at high signal-to-noise (corresponding to $M_{500} > 8 \times 10^{14} M_{\odot}$. Hence, any conclusion we may draw on the high radio halo fraction will be equally valid for later, more complete SZ catalogs.

Our results can be summarised as follows:
\begin{itemize}
\item{We have discovered diffuse radio emission in three clusters, namely Abell 1443, CIZAJ1938.3+5409, and RXCJ0949.8+1708. In RXCJ0949.8+1708 a tentative radio halo was already claimed by \citet{Venturi08}}
\item{Clusters that fall in the same region of the morphological diagrams $P_3/P_0,w$ and $c$ can or not host radio halo, and its power can follow or not the correlation, indicating that the dynamical status is not the only cause of radio emission.}
\item{The cluster Abell 1443 hosts peculiar emission which does not obviously fall into the categories of radio halo or radio relic. A bright $\Gamma$-shaped source is found at the cluster centre, which we interpret as a bright filamentary emission, similar to the one found in MACSJ0717+3745 \citep{Bonafede09b,vanweeren09}. Alternatively, it could be a peculiar wide-angle tail radio galaxy.}
\item{The cluster CIZAJ1938.3+5409 hosts diffuse emission that we classify as radio halo. The total power of the halo at 1.4 GHz, assuming a conservative $\alpha=1.2$ would be a factor eight below the  $P_{1.4\, {\rm GHz}}-$SZ correlations, placing the cluster in the ``radio quiet" zone.  
The cluster could be caught in its rapid ($\sim$ 0.1 -0.2 Gyr) transition phase between the $P_{1.4\, {\rm GHz}}$-SZ  correlation
and the upper-limit region of the $P_{1.4\, {\rm GHz}}$-SZ  plane, although no evident signs for an ongoing merger are visible in the X-ray image.\\
Alternatively, it could be the first direct detection of a hadronic radio halo. We have verified that this hypothesis is consistent with the most recent $\gamma$-ray upper limits by {\it Fermi} \citep{Ackermann14}. \\
Information about the spectral index will be helpful to discriminate between the two possibilities  or to point towards a different origin of the radio emission. 
It remains to be understood, though, if this kind or low power radio emission is common to all massive clusters or not.
In any case, this is the first direct detection of a cluster in its radio-off state. }
\item{The cluster RXCJ0949.8+1708 hosts diffuse emission on a Mpc scale that we classify as radio halo. The emission is elongated in the N-S direction and does not follow the X-ray emission from the gas. Although previous analysis based on optical and  X-ray observations indicate that the cluster is not in a merging stage, the X-ray morphological estimators suggest that the cluster is borderline between merger and non merger clusters.}
\item{No diffuse radio emission is detected in RXCJ1354.6+7715 at the sensitivity reached by our observations. Considering a halo size of $\sim$ 700 kpc and a detection theshold of 3$\sigma_{LR}$
we have placed an upper-limit to the radio emission which is a factor $\sim$ 5 below the $P_{1.4\, {\rm GHz}}$-SZ correlation.}

\end{itemize}
\section*{Acknowledgments}
The authors thank C. Jones for useful discussions. 
A.B.,  M.B, F.V, and F. dG acknowledge support by the research group FOR 1254 funded by the Deutsche Forschungsgemeinschaft:
``Magnetisation of interstellar and intergalactic media: the prospects of low-frequency radio observations".
We thank the staff of the GMRT that made these observation possible. GMRT is run by the National Centre for
Astrophysics of the Tata Institute of Fundamental Research. This research had made use of the NASA/IPAC Extragalactic Data Base 
(NED) which is operated by the JPL, California institute of technology under contract with the National Aeronautics and 
Space administration.

This publication makes use of data products from the Wide-field Infrared S urvey Explorer,  which is a joint project of the University of California, Los Angeles, and the Jet Propulsion Laboratory/California Institute of Technology, funded by the National Aeronautics and Space Administration

\bibliographystyle{mn2e}
\bibliography{master}

\begin{thebibliography}{}

\bibitem[\protect\citeauthoryear{{Ackermann}, {Ajello}, {Albert}, {Allafort},
  {Atwood}, {Baldini}, {Ballet}, {Barbiellini}, {Bastieri}, {Bechtol},
  {Bellazzini}, {Bloom}, {Bonamente}, {Bottacini} \& {Brandt}}{{Ackermann}
  et~al.}{2014}]{Ackermann14}
{Ackermann} M.,  {Ajello} M.,  {Albert} A.,  {Allafort} A.,  {Atwood} W.~B.,
  {Baldini} L.,  {Ballet} J.,  {Barbiellini} G.,  {Bastieri} D.,  {Bechtol} K.,
   {Bellazzini} R.,  {Bloom} E.~D.,  {Bonamente} E.,  {Bottacini} E.,
  {Brandt} T.~J.,  2014, \apj, 787, 18

\bibitem[\protect\citeauthoryear{{Basu}}{{Basu}}{2012}]{Basu12}
{Basu} K.,  2012, \mnras, 421, L112

\bibitem[\protect\citeauthoryear{{Blasi} \& {Colafrancesco}}{{Blasi} \&
  {Colafrancesco}}{1999}]{1999APh....12..169B}
{Blasi} P.,  {Colafrancesco} S.,  1999, Astroparticle Physics, 12, 169

\bibitem[\protect\citeauthoryear{{Bock}, {Large} \& {Sadler}}{{Bock}
  et~al.}{1999}]{SUMMS}
{Bock} D.~C.-J.,  {Large} M.~I.,    {Sadler} E.~M.,  1999, \aj, 117, 1578

\bibitem[\protect\citeauthoryear{{B{\"o}hringer}, {Pratt}, {Arnaud}, {Borgani},
  {Croston}, {Ponman}, {Ameglio}, {Temple} \& {Dolag}}{{B{\"o}hringer}
  et~al.}{2010}]{Boehringer10}
{B{\"o}hringer} H.,  {Pratt} G.~W.,  {Arnaud} M.,  {Borgani} S.,  {Croston}
  J.~H.,  {Ponman} T.~J.,  {Ameglio} S.,  {Temple} R.~F.,    {Dolag} K.,  2010,
  \aap, 514, A32

\bibitem[\protect\citeauthoryear{{B{\"o}hringer}, {Voges}, {Huchra}, {McLean},
  {Giacconi}, {Rosati}, {Burg}, {Mader}, {Schuecker}, {Simi{\c c}}, {Komossa},
  {Reiprich}, {Retzlaff} \& {Tr{\"u}mper}}{{B{\"o}hringer}
  et~al.}{2000}]{NORAS}
{B{\"o}hringer} H.,  {Voges} W.,  {Huchra} J.~P.,  {McLean} B.,  {Giacconi} R.,
   {Rosati} P.,  {Burg} R.,  {Mader} J.,  {Schuecker} P.,  {Simi{\c c}} D.,
  {Komossa} S.,  {Reiprich} T.~H.,  {Retzlaff} J.,    {Tr{\"u}mper} J.,  2000,
  \apjs, 129, 435

\bibitem[\protect\citeauthoryear{{Bonafede}, {Br{\"u}ggen}, {van Weeren},
  {Vazza}, {Giovannini}, {Ebeling}, {Edge}, {Hoeft} \& {Klein}}{{Bonafede}
  et~al.}{2012}]{Bonafede12}
{Bonafede} A.,  {Br{\"u}ggen} M.,  {van Weeren} R.,  {Vazza} F.,  {Giovannini}
  G.,  {Ebeling} H.,  {Edge} A.~C.,  {Hoeft} M.,    {Klein} U.,  2012, \mnras,
  426, 40

\bibitem[\protect\citeauthoryear{{Bonafede}, {Feretti}, {Giovannini}, {Govoni},
  {Murgia}, {Taylor}, {Ebeling}, {Allen}, {Gentile} \&
  {Pihlstr{\"o}m}}{{Bonafede} et~al.}{2009}]{Bonafede09b}
{Bonafede} A.,  {Feretti} L.,  {Giovannini} G.,  {Govoni} F.,  {Murgia} M.,
  {Taylor} G.~B.,  {Ebeling} H.,  {Allen} S.,  {Gentile} G.,    {Pihlstr{\"o}m}
  Y.,  2009, \aap, 503, 707

\bibitem[\protect\citeauthoryear{{Bonafede}, {Feretti}, {Murgia}, {Govoni},
  {Giovannini}, {Dallacasa}, {Dolag} \& {Taylor}}{{Bonafede}
  et~al.}{2010}]{Bonafede10}
{Bonafede} A.,  {Feretti} L.,  {Murgia} M.,  {Govoni} F.,  {Giovannini} G.,
  {Dallacasa} D.,  {Dolag} K.,    {Taylor} G.~B.,  2010, \aap, 513, A30

\bibitem[\protect\citeauthoryear{{Bonafede}, {Govoni}, {Feretti}, {Murgia},
  {Giovannini} \& {Br{\"u}ggen}}{{Bonafede} et~al.}{011a}]{Bonafede11a}
{Bonafede} A.,  {Govoni} F.,  {Feretti} L.,  {Murgia} M.,  {Giovannini} G.,
  {Br{\"u}ggen} M.,  {2011a}, \aap, 530, A24+

\bibitem[\protect\citeauthoryear{{Bonafede}, {Intema}, {Br{\"u}ggen},
  {Russell}, {Ogrean}, {Basu}, {Sommer}, {van Weeren}, {Cassano}, {Fabian} \&
  {R{\"o}ttgering}}{{Bonafede} et~al.}{2014}]{Bonafede14b}
{Bonafede} A.,  {Intema} H.~T.,  {Br{\"u}ggen} M.,  {Russell} H.~R.,  {Ogrean}
  G.,  {Basu} K.,  {Sommer} M.,  {van Weeren} R.~J.,  {Cassano} R.,  {Fabian}
  A.~C.,    {R{\"o}ttgering} H.~J.~A.,  2014, \mnras, 444, L44

\bibitem[\protect\citeauthoryear{{Bonafede}, {Vazza}, {Br{\"u}ggen}, {Murgia},
  {Govoni}, {Feretti}, {Giovannini} \& {Ogrean}}{{Bonafede}
  et~al.}{2013}]{Bonafede13}
{Bonafede} A.,  {Vazza} F.,  {Br{\"u}ggen} M.,  {Murgia} M.,  {Govoni} F.,
  {Feretti} L.,  {Giovannini} G.,    {Ogrean} G.,  2013, \mnras, 433, 3208

\bibitem[\protect\citeauthoryear{{Brown}, {Emerick}, {Rudnick} \&
  {Brunetti}}{{Brown} et~al.}{2011}]{Brown11}
{Brown} S.,  {Emerick} A.,  {Rudnick} L.,    {Brunetti} G.,  2011, \apjl, 740,
  L28

\bibitem[\protect\citeauthoryear{{Brunetti}, {Cassano}, {Dolag} \&
  {Setti}}{{Brunetti} et~al.}{2009}]{Brunetti09}
{Brunetti} G.,  {Cassano} R.,  {Dolag} K.,    {Setti} G.,  2009, \aap, 507, 661

\bibitem[\protect\citeauthoryear{{Brunetti}, {Giacintucci}, {Cassano}, {Lane},
  {Dallacasa}, {Venturi}, {Kassim}, {Setti}, {Cotton} \&
  {Markevitch}}{{Brunetti} et~al.}{2008}]{Brunetti08}
{Brunetti} G.,  {Giacintucci} S.,  {Cassano} R.,  {Lane} W.,  {Dallacasa} D.,
  {Venturi} T.,  {Kassim} N.~E.,  {Setti} G.,  {Cotton} W.~D.,    {Markevitch}
  M.,  2008, \nat, 455, 944

\bibitem[\protect\citeauthoryear{{Brunetti} \& {Jones}}{{Brunetti} \&
  {Jones}}{2014}]{BJ14}
{Brunetti} G.,  {Jones} T.~W.,  2014, ArXiv e-prints

\bibitem[\protect\citeauthoryear{{Brunetti} \& {Lazarian}}{{Brunetti} \&
  {Lazarian}}{2011}]{BL11}
{Brunetti} G.,  {Lazarian} A.,  2011, \mnras, 410, 127

\bibitem[\protect\citeauthoryear{{Brunetti}, {Setti}, {Feretti} \&
  {Giovannini}}{{Brunetti} et~al.}{2001}]{Brunetti01}
{Brunetti} G.,  {Setti} G.,  {Feretti} L.,    {Giovannini} G.,  2001, \mnras,
  320, 365

\bibitem[\protect\citeauthoryear{{Buote}}{{Buote}}{2001}]{Buote01}
{Buote} D.~A.,  2001, \apjl, 553, L15

\bibitem[\protect\citeauthoryear{{Buote} \& {Tsai}}{{Buote} \&
  {Tsai}}{1995}]{1995ApJ...452..522B}
{Buote} D.~A.,  {Tsai} J.~C.,  1995, \apj, 452, 522

\bibitem[\protect\citeauthoryear{{Cassano}, {Bernardi}, {Brunetti},
  {Br{\"u}ggen}, {Clarke}, {Dallacasa}, {Dolag}, {Ettori} \&
  {Giacintucci}}{{Cassano} et~al.}{2014}]{Cassano15}
{Cassano} R.,  {Bernardi} G.,  {Brunetti} G.,  {Br{\"u}ggen} M.,  {Clarke} T.,
  {Dallacasa} D.,  {Dolag} K.,  {Ettori} S.,    {Giacintucci} S.,  2014, in
  AASKA14 9 -13 June, 2014. Giardini Naxos, Italy. {Cluster Radio Halos at the
  crossroads between astrophysics and cosmology in the SKA era}.
p.~73

\bibitem[\protect\citeauthoryear{{Cassano}, {Ettori}, {Brunetti},
  {Giacintucci}, {Pratt}, {Venturi}, {Kale}, {Dolag} \& {Markevitch}}{{Cassano}
  et~al.}{2013}]{Cassano13}
{Cassano} R.,  {Ettori} S.,  {Brunetti} G.,  {Giacintucci} S.,  {Pratt} G.~W.,
  {Venturi} T.,  {Kale} R.,  {Dolag} K.,    {Markevitch} M.,  2013, \apj, 777,
  141

\bibitem[\protect\citeauthoryear{{Cassano}, {Ettori}, {Giacintucci},
  {Brunetti}, {Markevitch}, {Venturi} \& {Gitti}}{{Cassano}
  et~al.}{2010}]{Cassano10}
{Cassano} R.,  {Ettori} S.,  {Giacintucci} S.,  {Brunetti} G.,  {Markevitch}
  M.,  {Venturi} T.,    {Gitti} M.,  2010, \apjl, 721, L82

\bibitem[\protect\citeauthoryear{{Condon}, {Cotton}, {Greisen}, {Yin},
  {Perley}, {Taylor} \& {Broderick}}{{Condon} et~al.}{1998}]{NVSS}
{Condon} J.~J.,  {Cotton} W.~D.,  {Greisen} E.~W.,  {Yin} Q.~F.,  {Perley}
  R.~A.,  {Taylor} G.~B.,    {Broderick} J.~J.,  1998, \aj, 115, 1693

\bibitem[\protect\citeauthoryear{{Cotton}}{{Cotton}}{2008}]{OBIT}
{Cotton} W.~D.,  2008, \pasp, 120, 439

\bibitem[\protect\citeauthoryear{{Cuciti}, {Cassano}, {Brunetti}, {Dallacasa},
  {Kale}, {Ettori} \& {Venturi}}{{Cuciti} et~al.}{2015}]{Cuciti15}
{Cuciti} V.,  {Cassano} R.,  {Brunetti} G.,  {Dallacasa} D.,  {Kale} R.,
  {Ettori} S.,    {Venturi} T.,  2015, ArXiv e-prints

\bibitem[\protect\citeauthoryear{{Dennison}}{{Dennison}}{1980}]{1980ApJ...239L..93D}
{Dennison} B.,  1980, \apjl, 239, L93

\bibitem[\protect\citeauthoryear{{Dolag} \& {En{\ss}lin}}{{Dolag} \&
  {En{\ss}lin}}{2000}]{DolagEnsslin2000}
{Dolag} K.,  {En{\ss}lin} T.~A.,  2000, \aap, 362, 151

\bibitem[\protect\citeauthoryear{{Donnert}, {Dolag}, {Brunetti} \&
  {Cassano}}{{Donnert} et~al.}{2013}]{Donnert13}
{Donnert} J.,  {Dolag} K.,  {Brunetti} G.,    {Cassano} R.,  2013, \mnras, 429,
  3564

\bibitem[\protect\citeauthoryear{{Donnert}, {Dolag}, {Cassano} \&
  {Brunetti}}{{Donnert} et~al.}{2010}]{donn10}
{Donnert} J.,  {Dolag} K.,  {Cassano} R.,    {Brunetti} G.,  2010, \mnras, 407,
  1565

\bibitem[\protect\citeauthoryear{{Ebeling}, {Edge}, {Allen}, {Crawford},
  {Fabian} \& {Huchra}}{{Ebeling} et~al.}{2000}]{rosat_BCS}
{Ebeling} H.,  {Edge} A.~C.,  {Allen} S.~W.,  {Crawford} C.~S.,  {Fabian}
  A.~C.,    {Huchra} J.~P.,  2000, \mnras, 318, 333

\bibitem[\protect\citeauthoryear{{Ebeling}, {Edge}, {Mantz}, {Barrett},
  {Henry}, {Ma} \& {van Speybroeck}}{{Ebeling} et~al.}{2010}]{Ebeling10}
{Ebeling} H.,  {Edge} A.~C.,  {Mantz} A.,  {Barrett} E.,  {Henry} J.~P.,  {Ma}
  C.~J.,    {van Speybroeck} L.,  2010, \mnras, 407, 83

\bibitem[\protect\citeauthoryear{{Ebeling}, {Mullis} \& {Tully}}{{Ebeling}
  et~al.}{2002}]{CIZA}
{Ebeling} H.,  {Mullis} C.~R.,    {Tully} R.~B.,  2002, \apj, 580, 774

\bibitem[\protect\citeauthoryear{{En{\ss}lin}, {Pfrommer}, {Miniati} \&
  {Subramanian}}{{En{\ss}lin} et~al.}{2011}]{Ensslin11}
{En{\ss}lin} T.,  {Pfrommer} C.,  {Miniati} F.,    {Subramanian} K.,  2011,
  \aap, 527, A99

\bibitem[\protect\citeauthoryear{{Feretti}, {Giovannini}, {Govoni} \&
  {Murgia}}{{Feretti} et~al.}{2012}]{Feretti12}
{Feretti} L.,  {Giovannini} G.,  {Govoni} F.,    {Murgia} M.,  2012, \aar

\bibitem[\protect\citeauthoryear{{Giocoli}, {Moreno}, {Sheth} \&
  {Tormen}}{{Giocoli} et~al.}{2007}]{Giocoli07}
{Giocoli} C.,  {Moreno} J.,  {Sheth} R.~K.,    {Tormen} G.,  2007, \mnras, 376,
  977

\bibitem[\protect\citeauthoryear{{Giovannini}, {Feretti}, {Girardi}, {Govoni},
  {Murgia}, {Vacca} \& {Bagchi}}{{Giovannini}
  et~al.}{2011}]{2011A&A...530L...5G}
{Giovannini} G.,  {Feretti} L.,  {Girardi} M.,  {Govoni} F.,  {Murgia} M.,
  {Vacca} V.,    {Bagchi} J.,  2011, \aap, 530, L5

\bibitem[\protect\citeauthoryear{{Giovannini}, {Tordi} \&
  {Feretti}}{{Giovannini} et~al.}{1999}]{GTF99}
{Giovannini} G.,  {Tordi} M.,    {Feretti} L.,  1999, \na, 4, 141

\bibitem[\protect\citeauthoryear{{Govoni}, {Dolag}, {Murgia}, {Feretti},
  {Schindler}, {Giovannini}, {Boschin}, {Vacca} \& {Bonafede}}{{Govoni}
  et~al.}{2010}]{Govoni10}
{Govoni} F.,  {Dolag} K.,  {Murgia} M.,  {Feretti} L.,  {Schindler} S.,
  {Giovannini} G.,  {Boschin} W.,  {Vacca} V.,    {Bonafede} A.,  2010, \aap,
  522, A105+

\bibitem[\protect\citeauthoryear{{Horesh}, {Maoz}, {Ebeling}, {Seidel} \&
  {Bartelmann}}{{Horesh} et~al.}{2010}]{2010MNRAS.406.1318H}
{Horesh} A.,  {Maoz} D.,  {Ebeling} H.,  {Seidel} G.,    {Bartelmann} M.,
  2010, \mnras, 406, 1318

\bibitem[\protect\citeauthoryear{{Intema}, {van der Tol}, {Cotton}, {Cohen},
  {van Bemmel} \& {R{\"o}ttgering}}{{Intema} et~al.}{2009}]{SPAM}
{Intema} H.~T.,  {van der Tol} S.,  {Cotton} W.~D.,  {Cohen} A.~S.,  {van
  Bemmel} I.~M.,    {R{\"o}ttgering} H.~J.~A.,  2009, \aap, 501, 1185

\bibitem[\protect\citeauthoryear{{Intema}, {van Weeren}, {R{\"o}ttgering} \&
  {Lal}}{{Intema} et~al.}{2011}]{Intema11}
{Intema} H.~T.,  {van Weeren} R.~J.,  {R{\"o}ttgering} H.~J.~A.,    {Lal}
  D.~V.,  2011, \aap, 535, A38

\bibitem[\protect\citeauthoryear{{Kale}, {Venturi}, {Giacintucci}, {Dallacasa},
  {Cassano}, {Brunetti}, {Cuciti}, {Macario} \& {Athreya}}{{Kale}
  et~al.}{2015}]{Kale15}
{Kale} R.,  {Venturi} T.,  {Giacintucci} S.,  {Dallacasa} D.,  {Cassano} R.,
  {Brunetti} G.,  {Cuciti} V.,  {Macario} G.,    {Athreya} R.,  2015, \aap,
  579, A92

\bibitem[\protect\citeauthoryear{{Kelner}, {Aharonian} \& {Bugayov}}{{Kelner}
  et~al.}{2006}]{2006PhRvD..74c4018K}
{Kelner} S.~R.,  {Aharonian} F.~A.,    {Bugayov} V.~V.,  2006, \prd, 74, 034018

\bibitem[\protect\citeauthoryear{{Keshet} \& {Loeb}}{{Keshet} \&
  {Loeb}}{2010}]{2010ApJ...722..737K}
{Keshet} U.,  {Loeb} A.,  2010, \apj, 722, 737

\bibitem[\protect\citeauthoryear{{Khatri}}{{Khatri}}{2015}]{2015arXiv150500778K}
{Khatri} R.,  2015, ArXiv e-prints

\bibitem[\protect\citeauthoryear{{Lane}, {Cotton}, {van Velzen}, {Clarke},
  {Kassim}, {Helmboldt}, {Lazio} \& {Cohen}}{{Lane} et~al.}{2014}]{VLSSr}
{Lane} W.~M.,  {Cotton} W.~D.,  {van Velzen} S.,  {Clarke} T.~E.,  {Kassim}
  N.~E.,  {Helmboldt} J.~F.,  {Lazio} T.~J.~W.,    {Cohen} A.~S.,  2014,
  \mnras, 440, 327

\bibitem[\protect\citeauthoryear{{Liang}, {Hunstead}, {Birkinshaw} \&
  {Andreani}}{{Liang} et~al.}{2000}]{Liang2000}
{Liang} H.,  {Hunstead} R.~W.,  {Birkinshaw} M.,    {Andreani} P.,  2000, \apj,
  544, 686

\bibitem[\protect\citeauthoryear{{Maughan}, {Jones}, {Forman} \& {Van
  Speybroeck}}{{Maughan} et~al.}{2008}]{Maughan08}
{Maughan} B.~J.,  {Jones} C.,  {Forman} W.,    {Van Speybroeck} L.,  2008,
  \apjs, 174, 117

\bibitem[\protect\citeauthoryear{{Petrosian}}{{Petrosian}}{2001}]{Petrosian01}
{Petrosian} V.,  2001, \apj, 557, 560

\bibitem[\protect\citeauthoryear{{Pfrommer} \& {En{\ss}lin}}{{Pfrommer} \&
  {En{\ss}lin}}{2004a}]{pe04}
{Pfrommer} C.,  {En{\ss}lin} T.~A.,  2004a, \aap, 413, 17

\bibitem[\protect\citeauthoryear{{Pfrommer} \& {En{\ss}lin}}{{Pfrommer} \&
  {En{\ss}lin}}{2004b}]{2004A&A...426..777P}
{Pfrommer} C.,  {En{\ss}lin} T.~A.,  2004b, \aap, 426, 777

\bibitem[\protect\citeauthoryear{{Planck Collaboration}, {Ade}, {Aghanim},
  {Armitage-Caplan}, {Arnaud}, {Ashdown}, {Atrio-Barandela}, {Aumont},
  {Aussel}, {Baccigalupi} \& et al.}{{Planck Collaboration}
  et~al.}{2013}]{Planck13}
{Planck Collaboration} {Ade} P.~A.~R.,  {Aghanim} N.,  {Armitage-Caplan} C.,
  {Arnaud} M.,  {Ashdown} M.,  {Atrio-Barandela} F.,  {Aumont} J.,  {Aussel}
  H.,  {Baccigalupi} C.,    et al. 2013, ArXiv e-prints

\bibitem[\protect\citeauthoryear{{Planck Collaboration}, {Ade}, {Aghanim},
  {Arnaud}, {Ashdown}, {Aumont}, {Baccigalupi}, {Balbi}, {Banday}, {Barreiro}
  \& et al.}{{Planck Collaboration} et~al.}{2011}]{Planck11}
{Planck Collaboration} {Ade} P.~A.~R.,  {Aghanim} N.,  {Arnaud} M.,  {Ashdown}
  M.,  {Aumont} J.,  {Baccigalupi} C.,  {Balbi} A.,  {Banday} A.~J.,
  {Barreiro} R.~B.,    et al. 2011, \aap, 536, A8

\bibitem[\protect\citeauthoryear{{Rengelink}, {Tang}, {de Bruyn}, {Miley},
  {Bremer}, {Roettgering} \& {Bremer}}{{Rengelink} et~al.}{1997}]{wenss}
{Rengelink} R.~B.,  {Tang} Y.,  {de Bruyn} A.~G.,  {Miley} G.~K.,  {Bremer}
  M.~N.,  {Roettgering} H.~J.~A.,    {Bremer} M.~A.~R.,  1997, \aaps, 124, 259

\bibitem[\protect\citeauthoryear{{Santos}, {Rosati}, {Tozzi}, {B{\"o}hringer},
  {Ettori} \& {Bignamini}}{{Santos} et~al.}{2008}]{Santos08}
{Santos} J.~S.,  {Rosati} P.,  {Tozzi} P.,  {B{\"o}hringer} H.,  {Ettori} S.,
   {Bignamini} A.,  2008, \aap, 483, 35

\bibitem[\protect\citeauthoryear{{Scaife} \& {Heald}}{{Scaife} \&
  {Heald}}{2012}]{ScaifeHeald12}
{Scaife} A.~M.~M.,  {Heald} G.~H.,  2012, \mnras, 423, L30

\bibitem[\protect\citeauthoryear{{Sommer} \& {Basu}}{{Sommer} \&
  {Basu}}{2014}]{SommerBasu}
{Sommer} M.~W.,  {Basu} K.,  2014, \mnras, 437, 2163

\bibitem[\protect\citeauthoryear{{van Haarlem}, {Wise}, {Gunst}, {Heald},
  {McKean}, {Hessels}, {de Bruyn}, {Nijboer}, {Swinbank} \& {Fallows}}{{van
  Haarlem} et~al.}{2013}]{LOFAR}
{van Haarlem} M.~P.,  {Wise} M.~W.,  {Gunst} A.~W.,  {Heald} G.,  {McKean}
  J.~P.,  {Hessels} J.~W.~T.,  {de Bruyn} A.~G.,  {Nijboer} R.,  {Swinbank} J.,
     {Fallows} R. e.~a.,  2013, \aap, 556, A2

\bibitem[\protect\citeauthoryear{{van Weeren}, {R{\"o}ttgering}, {Bagchi},
  {Raychaudhury}, {Intema}, {Miniati}, {En{\ss}lin}, {Markevitch} \&
  {Erben}}{{van Weeren} et~al.}{2009}]{2009A&A...506.1083V}
{van Weeren} R.~J.,  {R{\"o}ttgering} H.~J.~A.,  {Bagchi} J.,  {Raychaudhury}
  S.,  {Intema} H.~T.,  {Miniati} F.,  {En{\ss}lin} T.~A.,  {Markevitch} M.,
  {Erben} T.,  2009, \aap, 506, 1083

\bibitem[\protect\citeauthoryear{{van Weeren}, {R{\"o}ttgering}, {Br{\"u}ggen}
  \& {Cohen}}{{van Weeren} et~al.}{2009}]{vanweeren09}
{van Weeren} R.~J.,  {R{\"o}ttgering} H.~J.~A.,  {Br{\"u}ggen} M.,    {Cohen}
  A.,  2009, \aap, 505, 991

\bibitem[\protect\citeauthoryear{{van Weeren}, {R{\"o}ttgering}, {Rafferty},
  {Pizzo}, {Bonafede}, {Br{\"u}ggen}, {Brunetti}, {Ferrari}, {Orr{\`u}},
  {Heald}, {McKean}, {Tasse}, {de Gasperin} \& {B{\^i}rzan}}{{van Weeren}
  et~al.}{2012}]{vanweeren12}
{van Weeren} R.~J.,  {R{\"o}ttgering} H.~J.~A.,  {Rafferty} D.~A.,  {Pizzo} R.,
   {Bonafede} A.,  {Br{\"u}ggen} M.,  {Brunetti} G.,  {Ferrari} C.,  {Orr{\`u}}
  E.,  {Heald} G.,  {McKean} J.~P.,  {Tasse} C.,  {de Gasperin} F.,
  {B{\^i}rzan} 2012, \aap, 543, A43

\bibitem[\protect\citeauthoryear{{Vazza}, {Eckert}, {Brueggen} \&
  {Huber}}{{Vazza} et~al.}{2015}]{Vazza15_p}
{Vazza} F.,  {Eckert} D.,  {Brueggen} M.,    {Huber} B.,  2015, ArXiv e-prints

\bibitem[\protect\citeauthoryear{{Venturi}, {Giacintucci}, {Dallacasa},
  {Cassano}, {Brunetti}, {Bardelli} \& {Setti}}{{Venturi}
  et~al.}{2008}]{Venturi08}
{Venturi} T.,  {Giacintucci} S.,  {Dallacasa} D.,  {Cassano} R.,  {Brunetti}
  G.,  {Bardelli} S.,    {Setti} G.,  2008, \aap, 484, 327

\bibitem[\protect\citeauthoryear{{Wright}, {Eisenhardt}, {Mainzer}, {Ressler},
  {Cutri}, {Jarrett}, {Kirkpatrick}, {Padgett}, {McMillan}, {Skrutskie},
  {Stanford}, {Cohen}, {Walker} \& {Mather}}{{Wright} et~al.}{2010}]{WISE}
{Wright} E.~L.,  {Eisenhardt} P.~R.~M.,  {Mainzer} A.~K.,  {Ressler} M.~E.,
  {Cutri} R.~M.,  {Jarrett} T.,  {Kirkpatrick} J.~D.,  {Padgett} D.,
  {McMillan} R.~S.,  {Skrutskie} M.,  {Stanford} S.~A.,  {Cohen} M.,  {Walker}
  R.~G.,    {Mather} J.~C. e.~a.,  2010, \aj, 140, 1868

\bibitem[\protect\citeauthoryear{{Zandanel}, {Pfrommer} \& {Prada}}{{Zandanel}
  et~al.}{2014}]{Zandanel14}
{Zandanel} F.,  {Pfrommer} C.,    {Prada} F.,  2014, \mnras, 438, 124

\end{thebibliography}

\label{lastpage}

\end{document}